%% file: ms.tex
\documentclass[12pt,preprint]{aastex}

\newcommand{\teff}{$T_{\mathrm{eff}}$}
\newcommand{\one}{HE 0054-2542}
\newcommand{\two}{HE 0507-1653}
\newcommand{\three}{HE 1005-1439}

\begin{document}

\title{CARBON ABUNDANCES OF THREE CARBON-ENHANCED METAL-POOR STARS FROM
       HIGH-RESOLUTION GEMINI-S/bHROS SPECTRA OF THE $\lambda 8727$ [\ion{C}{1}]
       LINE}
       
\author{Simon C. Schuler\altaffilmark{1,2}, Steven J. Margheim\altaffilmark{3}, 
        Thirupathi Sivarani\altaffilmark{4}, Martin Asplund\altaffilmark{5}, 
	Verne V. Smith\altaffilmark{1}, Katia Cunha\altaffilmark{1,6}, AND
	Timothy C. Beers\altaffilmark{4}}
	
\affil{
  \altaffiltext{1}{National Optical Astronomy Observatory/Cerro Tololo
                   Inter-American Observatory.  Casilla 603, La Serena, Chile:
		   sschuler@ctio.noao.edu, vsmith@noao.edu, cunha@noao.edu}
  \altaffiltext{2}{Leo Goldberg Fellow}
  \altaffiltext{3}{Gemini Observatory, Casilla 603, La Serena, Chile:
                   smargheim@gemini.edu}
  \altaffiltext{4}{Department of Physics and Astronomy, Center for the Study of
                   Cosmic Evolution, and Joint Institute for Nuclear
		   Astrophysics, Michigan State University, East Lansing, MI:
		   thirupathi@pa.msu.edu, beers@pa.msu.edu}
  \altaffiltext{5}{Max Planck Institute for Astrophysics, Postfach 1317, 85741,
                   Garching, Germany: asplund@mpa-garching.mpg.de}
  \altaffiltext{6}{On leave from Observat{\'o}rio Nacional, Rio de Janeiro,
                   Brazil}
  }

\begin{abstract}
We present the results from an analysis of the $\lambda 8727$ forbidden 
[\ion{C}{1}] line in high-resolution Gemini-S/bHROS spectra of three 
Carbon-Enhanced Metal-Poor (CEMP) stars.  Previous derivations of C abundances 
in CEMP stars primarily have used the blue bands of CH and the C$_2$ Swan 
system, features which are suspected to be sensitive to photospheric temperature 
inhomogeneities (the so-called 3D effects).  We find the [C/Fe] ratios based on 
the [\ion{C}{1}] abundances of the two most Fe-rich stars in our sample
(\two: $\mathrm{[Fe/H]} = -1.42$ and \one: $\mathrm{[Fe/H]} = -2.66$) 
to be in good agreement with previously determined values. For the most 
Fe-deficient star in our sample (\three: $\mathrm{[Fe/H]} = -3.08$), however, 
the [C/Fe] ratio is  found to be 0.34 dex lower than the published 
molecular-based value.  We have carried out 3D local thermodynamic equilibrium 
(LTE) calculations for [\ion{C}{1}], and the resulting corrections are found to be 
modest for all three stars, suggesting that the discrepancy between the [\ion{C}{1}] 
and molecular-based C abundances of \three\ is due to more severe 3D effects 
on the molecular lines.  Carbon abundances are also derived from \ion{C}{1} 
high-excitation lines and are found to be 0.45 -- 0.64 dex higher than the 
[\ion{C}{1}]-based abundances.  Previously published non-LTE (NLTE) \ion{C}{1} 
abundance corrections bring the [\ion{C}{1}] and \ion{C}{1} abundances into better 
agreement; however, targeted NLTE calculations for CEMP stars are clearly needed.  
We have also derived the abundances of nitrogen, potassium, and iron for each 
star.  The Fe abundances agree well with previously derived values, and the K 
abundances are similar to those of C-normal metal-poor stars. Nitrogen 
abundances have been derived from resolved lines of the CN red system assuming 
the C abundances derived from the [\ion{C}{1}] feature. The abundances are 
found to be approximately 0.44 dex larger than literature values, which have 
been derived from CN blue bands near 3880 and 4215 {\AA}. We discuss evidence 
that suggests that analyses of the CN blue system bands underestimate the N 
abundances of metal-poor giants.
\end{abstract}


\section{INTRODUCTION}
The discovery that a significant fraction of very metal-poor (VMP; 
$\mathrm{[Fe/H]} \leq -2.0$) stars have highly enhanced abundances of carbon 
($\mathrm{[C/Fe]} \geq +1.0$) has spurred vigorous research efforts focused on 
delineating the nucleosynthetic histories of these objects and their role in 
the chemical evolution of the Galaxy.  The actual fraction of VMP stars that 
are carbon-enhanced has yet to be precisely determined, but current estimates 
range from $\sim 10\%$ to $\sim 25\%$ 
\citep{2005ApJ...633L.109C,2005NuPhA.758..312M,2006ApJ...652.1585F,2006ApJ...652L..37L}.  
This fraction rises at lower metallicities, reaching $\sim 40\%$ of stars with 
$\mathrm{[Fe/H]} \leq -3.5$, and stars with $\mathrm{[Fe/H]} < -4.0$, only three 
of which are currently known, are all carbon-enhanced.  The increasing incidence 
of carbon enhancement at lower and lower metallicities implies that the 
nucleosynthetic pathway(s) leading to these Carbon-Enhanced Metal-Poor (CEMP) 
stars is highly efficient at low metallicities and that it played an important role 
in the nucleosynthetic history of the early Galaxy.  Indeed, the fraction of CEMP 
stars as a function of metallicity may have critical implications for the initial mass 
function (IMF) in the early Universe \citep{2007ApJ...665.1361T}.

The manner in which the majority of VMP stars are discovered, and how CEMP 
stars are subsequently identified, follows a well established 
procedure\footnotemark[6].  Stars are first tagged as VMP candidates based on 
the strength of the \ion{Ca}{2} K line in low-resolution spectra from 
objective-prism surveys, in particular the HK survey (Beers, Preston, \& 
Shectman 1985, 1992; Beers 1999) and the Hamburg/ESO survey 
\citep[HES;][]{2000A&A...358...77W,2003RvMA...16..191C}.  Medium-resolution spectra 
of the VMP candidates are then obtained and used to identify {\it bona fide} VMP 
stars within the sample.  The medium-resolution spectrum of a given VMP star is used 
to estimate its metallicity ([Fe/H]) by analyzing the strength of the \ion{Ca}{2} 
K line as a function of the broadband colors of the star, and inspection of the CH 
G-band found at 4300 {\AA} reveals whether or not the star is enhanced in C 
\citep[e.g.,][]{2005AJ....130.2804R}.  Estimates of a star's C abundance can be obtained 
from the CH G-band; however, in the spectra of many CEMP stars, the G-band feature is 
sufficiently strong as to be saturated, rendering it incapable of providing 
accurate abundance determinations.  In these cases, C$_2$ lines of the Swan 
system can be used instead to estimate C abundances.  Once identified, VMP 
stars are often slated for high-resolution spectroscopic follow-up studies, 
with which more accurate abundances of Fe, C, and numerous other elements, as 
well as important ratios such as $^{12}$C/$^{13}$C, can be derived.

\footnotetext[6]{Here we describe the identification of metal-poor stars based 
on the low-resolution spectra of objective-prism surveys, the most prolific 
sources of VMP stars.  For a description of other methods used to find 
metal-poor stars, please see \citet{2005ARA&A..43..531B}.}

Analyses of C abundances in high-resolution spectroscopic follow-up studies of 
CEMP stars focus on the same molecular features as those analyzed in 
medium-resolution spectra, namely the blue bands of the CH and C$_2$ Swan
systems, because it is the molecular features that stay strong and most easily 
measurable in the spectra of VMP giants.  Another reason is that atomic 
\ion{C}{1} lines result from high-excitation transitions, and if measurable,
these lines are not believed to be accurate abundance indicators because of 
their sensitivity to non-local thermodynamic equilibrium (NLTE) effects
\citep{2005ARA&A..43..481A,2006A&A...458..899F}.  However, the abundances 
derived from the CH and C$_2$ features in the spectra of VMP stars using 
standard 1-dimensional (1D) local thermodynamic equilibrium (LTE) analyses may 
not be accurate either.  Abundances derived from molecular features are 
generally very sensitive to the temperature structure of stellar atmosphere 
models, and studies of time-dependent 3-dimensional (3D) hydrodynamical models 
have shown that the molecular lines are susceptible to temperature 
inhomogeneities due to photospheric granulation, the so-called 3D effects 
\citep{2005ARA&A..43..481A}.  The effects on molecular lines are such that 
abundances based on 3D models are lower than those derived using 1D models, 
with differences as large as $\sim 1.0$ dex at $\mathrm{[Fe/H]} = -3$ (Asplund \& 
Garc{\'i}a P{\'e}rez 2001; Collet, Asplund, \& Trampedach 2007).

The derivation of accurate C abundances of CEMP stars is essential to their 
proper characterization and subsequently to the correct interpretation of their 
role in the chemical evolution of the Galaxy.  Here we present the results of 
our investigation into the accuracy of published C abundances as derived from 
CH and C$_2$ features by analyzing the forbidden [\ion{C}{1}] line at 8727.13 
{\AA} in high-resolution Gemini-S/bHROS spectra of three CEMP stars: 
\object{HE 0054-2542} (\object{CS 22942-019}), \object{HE 0507-1653}, and 
\object{HE 1005-1439}.  The $\lambda 8727$ [\ion{C}{1}] line is known to be 
immune to NLTE effects and to be a highly reliable abundance indicator 
\citep{1999A&A...342..426G}, and it has been used to derive the C abundance of
the Sun (e.g., Lambert 1978; Allende Prieto, Lambert, \& Asplund 2002; Asplund
et al. 2005b), field dwarfs \citep[e.g.,][]{1999A&A...342..426G,2006MNRAS.367.1181B}, 
R Coronae Borealis stars \citep[e.g.,][]{2004MNRAS.353..143P}, and Cepheid
variables \citep[e.g.,][]{1981ApJ...245.1018L}.  For CEMP stars, it provides an 
excellent benchmark for C abundances derived from molecular features.

\section{ASTROPHYSICAL SPECTROSCOPY OF CARBON}
Carbon abundances can be derived from both atomic and molecular features in the
spectra of late-type stars.  The available atomic lines include numerous 
high-excitation ($\chi \geq 7.48 \; \mathrm{eV}$) \ion{C}{1} lines found 
throughout the optical and near-IR spectral regions and a single [\ion{C}{1}] 
forbidden feature at 8727.13 {\AA}.  Suitable molecular features include 
electronic and rotational-vibrational transitions of CH and C$_2$, which are 
also found throughout the optical and near-IR spectral regions.  Deriving C 
abundances from CO and CN transitions is also possible, but prior knowledge 
of the stellar O and N abundances is needed..  For CEMP stars, it is more 
often the case that the C abundance of a star is known from an analysis of 
CH or C$_2$ features, and the CO and CN lines are used for the derivation of O 
and N abundances.  However, for cool environments such as those found in the 
atmospheres of red giants, a significant fraction of C is tied up in CO, so C 
abundances derived from \ion{C}{1}, [\ion{C}{1}], CH and C$_2$ lines can also
be sensitive to the adopted O abundance.  Below we discuss the two sets of atomic 
and molecular C features in the context of using them to derive accurate 
abundances, particularly of CEMP stars.

\subsection{\ion{C}{1}}
The permitted neutral C lines in the spectra of late-type stars all originate 
from high-excitation transitions, and they all are sensitive to NLTE effects.  
The NLTE effects are such that C abundances are overestimated by LTE analyses 
(lines are weaker in LTE compared to NLTE), and thus NLTE corrections are 
negative.  Detailed accounts of modeling the C atom and the nature of NLTE 
effects are given by \citet{2005ARA&A..43..481A} and 
\citet{2006A&A...458..899F}.  

The stellar parameter space for which NLTE effects and the high-excitation 
\ion{C}{1} lines has been investigated is rather limited, but in general the 
published studies agree quite well.  For solar-like stars (stars with similar 
\teff, $\log \; g$, and metallicity as the Sun), NLTE corrections are typically 
-0.05 to -0.1 dex 
\citep{1990A&A...237..125S,2005ARA&A..43..481A,2005PASJ...57...65T,2006A&A...458..899F}.
These corrections are predicted to increase in magnitude, i.e., become more negative, 
with increasing \teff\ and decreasing $\log g$, as well as with increasing line 
strength.  Metallicity apparently has a minor effect on predicted NLTE corrections, 
with the peak correction occurring near $\mathrm{[Fe/H]} = -1.0$ and then tapering off 
slightly by no more than 0.10 dex at $\mathrm{[Fe/H]} = -3.0$. 
Interestingly, \citet{2006A&A...458..899F} note that the assumed [C/Fe] ratio
also affects the predicted NLTE corrections such that the magnitude of the 
corrections become larger for increasing [C/Fe] ratios since the line-formation region
is shifted outwards in the atmospheres where the departures from LTE is more
pronounced. One would thus expect this effect to be significant in the analysis of 
CEMP stars.  

\subsection{[\ion{C}{1}]}
The [\ion{C}{1}] forbidden line results from a low-excitation ($\chi = 1.26 \; 
\mathrm{eV}$) electric quadrapole ($D^{1}_{2}$ -- $S^{1}_{0}$) transition in 
carbon's ground state configuration ($2s^{2}2p^{2}$).  The ground state of 
\ion{C}{1} is populated according to Boltzmann statistics, and thus transitions 
from this state are formed in LTE.  The next three energy levels, of which the 
[\ion{C}{1}] transition arises from the first,  are strongly coupled 
through collisions to the ground state, assuring that they too are in LTE 
\citep{1990A&A...237..125S}.  For this reason, the [\ion{C}{1}] line is 
expected to be a highly accurate abundance indicator and not susceptible to 
NLTE effects \citep[e.g.,][]{1999A&A...342..426G}.

Deriving accurate C abundances from the [\ion{C}{1}] line is not without its 
challenges, however.  \citet{1967SoPh....2...34L} preliminarily identified a 
possible blending \ion{Fe}{1} line at 8727.10 {\AA}, but along with subsequent authors 
\citep[e.g.][]{1999A&A...342..426G,2002ApJ...573L.137A,2005A&A...431..693A},
they determined that the blending \ion{Fe}{1} feature contributes negligibly to the 
[\ion{C}{1}] line strength in the solar spectrum.  However, for completeness, 
the blending feature should be included in the $\lambda 8727$ linelist, 
particularly for stars with [C/Fe]$<0$.  Also, in the solar spectrum there is a 
strong ($\sim 100$ m{\AA}) \ion{Si}{1} line at 8728.01 {\AA}, just 0.88 {\AA} 
to the red of the [\ion{C}{1}] feature.  The \ion{Si}{1} line can affect the continuum 
in the $\lambda 8727$ region, making high-resolution spectroscopy essential to 
the accurate measurement of the [\ion{C}{1}] line.  Neither of these concerns is 
applicable to metal-poor stars, for which the blending \ion{Fe}{1} and neighboring 
\ion{Si}{1} features become increasingly weak at sub-solar metallicities.  Unfortunately, 
the [\ion{C}{1}] line also becomes weak at low metallicities, vanishingly so at 
$\mathrm{[Fe/H]} < -1.0$ \citep{2004A&A...414..931A}, {\it except} for CEMP 
stars.  As we demonstrate below, the [\ion{C}{1}] line can remain strong enough 
for reliable measurements in the spectra of CEMP stars down to at least 
$\mathrm{[Fe/H]} \sim -3.0$.

As a final note, we point out that temperature inhomogeneities due to 
photospheric granulation, the so-called 3D effects, do not greatly affect 
abundances derived from neutral atomic C lines \citep{2005ARA&A..43..481A}.  
The difference in the solar C abundance as derived from the [\ion{C}{1}] line 
using 3D hydrodynamical LTE models and conventional 1D LTE models amounts to 
$\la 0.05$ dex \citep{2005A&A...431..693A}.  While the susceptibility of the 
[\ion{C}{1}] line to LTE 3D effects as a function of metallicity has not been 
investigated, the high-excitation \ion{C}{1} lines are expected to remain 
relatively impervious to them at all metallicities due to their large depths of 
formation \citep{2005ARA&A..43..481A}.

\subsection{CH and C$_2$}
These molecular features are likely relatively immune to NLTE 
effects, although little computational work has been done in this area 
\citep{2005ARA&A..43..481A}.  What is clear is that the molecular lines are 
highly sensitive to temperature inhomogeneities and by extension, 3D effects.  
For the Sun, 3D corrections for abundances derived from CH and C$_2$ lines 
have been estimated to range from 0.00 to -0.15 dex  (1D abundances are higher), 
depending on the lines being analyzed \citep{2005A&A...431..693A}.  The 
divergence between C abundances derived from these molecular lines using 1D and 
3D models grows with decreasing metallicity, reaching differences ranging from 
-0.5 to -0.8 dex for giants and sub-giants at $\mathrm{[Fe/H]} = -3$ 
\citep{2005ARA&A..43..481A,2007A&A...469..687C}.  

The susceptibility of the CH and C$_2$ features to 3D effects is unfortunate, 
because it is these lines that generally remain strong at low metallicities.  
Indeed, the CH band near 4300 {\AA} and lines of the C$_2$ Swan system 
at 4730, 5170, and 5635 {\AA} are the features most often used to derive C 
abundances of CEMP stars.  Many authors have noted that the C abundances 
derived from these molecular features may be in error by as much as 0.60 dex or 
more \citep[e.g.,][]{2007ApJ...655..492A,2007ApJ...658..534F}, but because 3D 
hydrodynamical models tailored for CEMP stars have not been constructed, 
accurate corrections to 1D abundances cannot be made.  This is the source of the 
motivation for the research presented in this paper.  By deriving abundances 
from an accurate abundance indicator, the [\ion{C}{1}] forbidden line, we can 
provide some insight into the relative accuracy of the C abundances derived from 
molecular lines in the spectra of CEMP stars.

\section{OBSERVATIONS AND DATA REDUCTION}
\subsection{Observations}
Multiple spectra of three CEMP stars were taken on UT 2006 December 26 and 27 
at the Gemini-S telescope with bHROS, the bench-mounted High-Resolution Optical 
Spectrograph.  The spectrograph is located within a thermal enclosure in the 
pier lab of the Gemini-S telescope and is fiber-fed from the telescope's 
Cassegrain focus via a fiber cassette inserted into the Gemini Multi-Object 
Spectrograph (GMOS-South).  The object-only mode of operation was used for our
observations; this mode utilizes a single $0.9\arcsec$ fiber that feeds a 
dedicated image slicer to produce a `slit' that is $0.14\arcsec$ wide and 
$\sim6.5\arcsec$ long, projected to the camera focal plane.  A slicer rotation 
mechanism is used to ensure a `vertical' slit at the observed spectral 
wavelength.  The spectrograph is cross-dispersed using a set of fused silica 
prisms to ensure the echelle orders are well separated on the detector, a 
single 2048x4608 E2V CCD with 13.5$\mu$m pixels.  The wavelength range over 
which bHROS is operational spans from 4000 to 10,000 {\AA}.

For our observations, $4 \times 2$ (spatial direction $\times$ dispersion
direction) binning was used to reduce the spectrograph's working resolution of 
$R = \lambda / \delta \lambda = 150,000$ to an effective resolution of $R = 
75,000$ in order to increase the signal-to-noise (S/N) ratio of the spectra and 
to lessen the impact of read noise, which can be significant given the 
detector's intrinsically high read noise of 5.3 e$^-$ and wide spectral orders 
(200 unbinned pixels).  The grating was configured to produce a central 
wavelength on the detector of 8352 {\AA}.  In this configuration, incomplete 
coverage from 7135 -- 9170 {\AA} was obtained over seven echelle orders.  The 
continuous coverage available within a single order is about 88 {\AA}, 
depending on the order.    

The spectra were obtained in queue operations during photometric conditions 
with optical seeing varying between 0.35 and $0.8\arcsec$, with the majority of 
the delivered image quality better than $0.5\arcsec$.  The total exposure time 
was 9880s for \one\ (3 exposures), 6280s for \two\ (2 exposures), and 15900s 
for \three\ (5 exposures); the S/N ratios of the combined spectra at 8727 {\AA} 
are 114 for \one, 149 for \two, and 65 for \three.  A detailed observing log is 
presented in Table 1.  In addition to our target spectra, we also obtained a 
calibration set of daily biases, flats, and ThAr arc spectra. 

\subsection{Reductions}
Processing of the spectra was completed following normal echelle reduction 
practices with standard IRAF packages, with careful adaptations to suit the 
bHROS data.  The raw spectra were bias corrected by overscan and bias image 
subtraction.  The inter-order scattered light is typically $\sim$5 ADU and 
varies by only a few tenths of an ADU over 50 pixels, which is the width of our 
binned orders and much larger than the spectral features we are observing.  
Since the uncertainty in the smoothed fit to the scattered light is larger than 
the gradient in the scattered light, no scattered light correction was 
performed.  Cosmic rays were removed using L.A. Cosmic 
\citep{2001PASP..113.1420V}.  The images were then flat-fielded to remove both 
the pixel-to-pixel variations and fringing from our spectra.   The E2V CCD 
used in bHROS suffers from severe fringing in the red.  However, as the light 
path for the flat observations is identical to our target observations, the 
illumination pattern on the detector is constant, and the fringe pattern is 
consistent in all spectra.  Thus, the fringe patterns in our spectra are very 
well removed in the flat fielding process.  
	
The use of a prism cross-disperser combined with a long `slit length' causes 
the spectral orders to be tilted severely away from the central order,
potentially resulting in the loss of resolution during the extraction process. 
To reduce the impact of this tilt on the resolution of our extracted spectra, 
each of the seven spectral orders was divided into 14 sub-apertures (7 orders 
$\times$ 14 sub-apertures $=$ 98 total sub-apertures), and then each 
sub-aperture was extracted individually using optimal extraction methods.  The 
extracted spectra were then blaze corrected using the extracted flat field 
spectra.  The individual orders of the ThAr spectra were also divided into 14
sub-apertures, and wavelength solutions were derived for each of the 98 
sub-apertures.  The typical RMS scatter of the dispersion solutions was 0.0025 
pixels.  The wavelength solutions and heliocentric velocity corrections were 
applied to the individual sub-apertures, which were then recombined into their 
respective orders, achieving a fully reduced 1-D echelle spectrum composed of 
seven spectral orders.  Finally, the spectra were continuum normalized to 
produce our final spectra for analysis.

\section{ANALYSIS}
\subsection{Stellar Parameters}
Determining the physical parameters-\teff\, surface gravity ($\log g$), and 
microturbulent velocity ($\xi$)- of a stellar sample is always a critical 
component of any chemical abundance analysis.  In general, this is not a 
trivial exercise and requires spectra of sufficient quality and wavelength 
coverage to allow for the derivation of the parameters directly from the spectra 
themselves, and/or accurate photometry, color-\teff\ relations, and isochrones 
with which the parameters can be determined indirectly.  For metal-poor stars, 
the spectroscopic method may not be an option at all due to the paucity of the 
necessary metal lines in their spectra.  This is the case for our sample, and 
thus we have adopted stellar parameters appearing in the recent literature.  
This has the added advantage of allowing a more direct comparison of our 
[C I]-based abundances to those derived previously.  The adopted parameters for 
each star are given in Table 2 and are described here:

{\bf HE 0054-2542:} This star, also known as CS 22942-019, has been studied by
\citet{2001AJ....122.1545P} and \citet{2002ApJ...580.1149A}.  Preston \& Sneden
used the Revised Yale Isochrones \citep{1987ryil.book.....G} and the observed
dereddened (\bv)$_0$ colors, corrected for the blanketing effects of the strong 
molecular C bands in the B and V filter passbands, of their stellar sample to 
obtain initial values of \teff\ and $\log g$.  They then proceeded to refine the
stellar parameters, including the microturbulent velocity, spectroscopically by
requiring the abundances of \ion{Fe}{1}, \ion{Fe}{2}, and \ion{Ti}{2} derived
using high-resolution spectra of their sample to be independent on a 
line-by-line basis of line strength, excitation potential, and ionization 
state.  This process was hampered by the modest S/N ratio and resolution of 
their spectra (typically S/N$\sim 30$, $R \sim 20,000$), and any adjustments to 
the initial parameter values were constrained to fall within the estimated 
parameter uncertainties.  The final parameters adopted by them for \one\ are 
$T_{\mathrm{eff}} = 4900 \; \mathrm{K}$, $\log g = 1.8$ (cgs), $\xi = 2.0 \; 
\mathrm{km \; s}^{-1}$, and $\mathrm{[Fe/H]} = -2.67$.

\citet{2002ApJ...580.1149A} used broadband photometry and a temperature scale
based on dereddened (\bv)$_0$ colors devised for carbon-rich metal-poor stars 
by \citet{2002ApJ...567.1166A} to estimate \teff\ for their stellar sample.  
Surface gravities and microturbulent velocities were then derived 
spectroscopically using high-resolution spectra by forcing ionization balance 
between the abundances of \ion{Fe}{1} and \ion{Fe}{2} and demanding that 
line-by-line \ion{Fe}{1} abundances were independent of equivalent width (EW).  
The resulting parameters found by \citet{2002ApJ...580.1149A} for \one\ are 
$T_{\mathrm{eff}} = 5000 \; \mathrm{K}$, $\log g = 2.4$ (cgs), $\xi = 2.1 \; 
\mathrm{km \; s}^{-1}$, and $\mathrm{[Fe/H]} = -2.64$.

The agreement in the stellar parameters between the \citet{2001AJ....122.1545P} 
and \citet{2002ApJ...580.1149A} studies is quite good, with the possible
exception of the surface gravity.  Both groups derived $\log g$ for this star
spectroscopically, that is to say by forcing ionization balance; however, it is 
likely that the low signal-to-noise ratio and moderate resolution spectra of 
Preston \& Sneden adversely affected their parameter derivations.  Indeed, the 
authors point out that the line-to-line scatter in the individual abundances 
are typically 0.2 -- 0.3 dex, which prevented them from doing a detailed 
assessment of their stellar parameters.  Thus, we have adopted the $\log g$ 
value for \one\ derived by \citet{2002ApJ...580.1149A}, who used higher quality 
spectra in their analysis ($R = 50,000$ and S/N $\sim 60$).  Likewise, we have 
adopted the \teff\ and [Fe/H] abundance of \citet{2002ApJ...580.1149A}, along 
with $\xi = 2.0 \; \mathrm{km \; s}^{-1}$, for this star.

{\bf HE 0507-1653:} This star appears in a single published study, that of
\citet{2007ApJ...655..492A}.  Effective temperatures for their sample of stars
were estimated using broadband photometry and the color-\teff\ relations of
\citet{1999A&AS..140..261A}.  Although \teff\ values were calculated using 
($V - K$), ($V - R$), ($V - I$), and ($R - I$) colors, the ($V - K$)-based 
\teff\ were adopted for most of their sample, because ($V - K$) has been 
demonstrated to provide the best photometric \teff\ estimate for CEMP stars 
\citep{2002AJ....124..470C}.  The \teff\ were then refined, and $\log g$, $\xi$,
and the Fe abundances were derived, spectroscopically.  For \two, however, too 
few weak \ion{Fe}{1} lines were present in its spectrum to allow for an accurate
microturbulent velocity to be determined, so $\xi = 2.0 \; \mathrm{km \; 
s}^{-1}$, a value close to the average for the remaining stars in their sample, 
was assumed.  The final stellar parameters for \two\ from 
\citet{2007ApJ...655..492A} and those adopted here are $T_{\mathrm{eff}} = 5000 
\; \mathrm{K}$, $\log g = 2.4$, $\xi = 2.0 \; \mathrm{km \; s}^{-1}$, and 
$\mathrm{[Fe/H]} = -1.38$.

{\bf HE 1005-1439:} \citet{2007ApJ...655..492A} is also the source for the
stellar parameters adopted for this star.  The ($V - K$)-based \teff\ for 
\three, though, was found to be $\sim 650 \; \mathrm{K}$ higher than the 
average \teff\ calculated using the ($V - R$), ($V - I$), and ($R - I$) colors, 
and the authors adopted the lower \teff, suspecting the ($V - K$)-based \teff\ 
was in error.  We too adopt this lower temperature, \teff $= 5000 \; 
\mathrm{K}$, as well as the other parameters from the Aoki et al. study: $\log 
g = 1.9$, $\xi = 2.0 \; \mathrm{km \; s}^{-1}$, and $\mathrm{[Fe/H]} = -3.17$.

\subsection{Model Atmospheres}
The structure of a stellar photosphere is dependent in part on the star's 
chemical composition due to the contribution of metals to the continuous 
opacity.  As the concentration of metals increases, the electron pressure
increases as a result of a greater number of free electrons, which in turn 
affects the temperature and pressure stratification of the photosphere.  
Metals such as C, Si, Al, Mg, and Fe contribute the most to the continuous 
opacity, and so the photospheres of CEMP stars, with their enhanced C 
abundances, are expected to have temperature and pressure profiles that differ 
from those of ``C-normal'' metal-poor stars.  In light of this, using model 
stellar atmospheres constructed with scaled solar abundances for a chemical 
abundance analysis of CEMP stars is less than optimal, as the change in the 
temperature and pressure structure in the line-forming regions due to the 
enhanced C, if not properly modeled, could result in inaccurate abundance 
derivations.

Unfortunately, C-enhanced models are not as readily available as models with 
scaled solar abundances, so it is not surprising that the majority of CEMP star 
abundance studies found in the literature have made use of the latter.  
Nonetheless, a handful of these studies have addressed the use of C-enhanced 
models versus solar-scaled models \citep[e.g.,][]{2003AJ....125..875L,2007ApJ...655..492A},
and one study, that of \citet{2000A&A...353..557H}, has provided results 
from a more comprehensive investigation.  Hill et al. compared the profiles of 
the temperature and the molecular partial pressures of three C-bearing molecules 
versus the gas pressure of a C-enhanced model atmosphere based on the revised 
MARCS code \citep{1992A&A...256..551P} and two solar-scaled models, one from the 
grids of Kurucz \citep{kurucz} and one from the MARCS code 
\citep{1975A&A....42..407G}, with parameters \teff$= 4500 \; \mathrm{K}$, $log 
\; g = 1$, and $[\mathrm{Fe/H}] = -3$.  They found that the temperature gradient 
in the C-enhanced model differed significantly from those of the solar-scaled 
models.  The differences in the derived abundances using the different models 
are not quantified by the authors, but they do show that the upper layers of the 
C-enhanced atmosphere, where strong lines generally form, are cooler than in the 
C-normal atmospheres.  Conversely, in the deeper layers where intermediate and
weaker lines form, they showed that the C-enhanced models are hotter.  These 
differences in the temperature structure can affect abundance derivations, 
especially those based on molecular features which are highly sensitive to 
temperature.

While a more rigorous study of metal-poor stellar atmosphere models enriched 
with C, as well as N and O, and their effect on stellar abundance derivations 
is needed, we have taken a more empirical approach to addressing the issue.  
We have generated both solar scaled and C-enhanced models using the adopted 
stellar parameters (Table 2) for each star in our sample and have used both
models in our abundance analysis.  The solar scaled models have been 
interpolated from the ATLAS9 grids of Kurucz\footnotemark[7].  These models 
assume local thermodynamic equilibrium (LTE) and use the convective overshoot 
approximation.  The C-enhanced models have been generated using the LTE stellar 
atmosphere code ATLAS12 \citep{kurucz96}.  These models also include 
enhancements in N and O.  The C and N abundances, which are given in Table 2, 
are from the literature: \citet{2002ApJ...580.1149A} for \one, and 
\citet{2007ApJ...655..492A} for \two\ and \three.  Oxygen abundances are not 
available for these stars, so we adopted an abundance of $\mathrm{[O/Fe]} = 
+0.40$, which is typical for Galactic halo stars. 

\footnotetext[7]{See http://kurucz.harvard.edu/grids.html}

\subsection{Abundance Derivations}
The abundance analysis of the three stars in our sample has been carried out
using the stellar line analysis code MOOG \citep{1973ApJ...184..839S}.  Spectral
synthesis was used to fit the [\ion{C}{1}] line with the C abundance being the
only free parameter.  The oscillator strength for this transition, $\log gf =
-8.136$, is taken from \citet{2002ApJ...573L.137A}, who averaged the values
given by \citet{1997A&AS..123..159G} and \citet{1993A&AS...99..179H}.  This
value is in agreement with that provided by the NIST database\footnotemark[8] 
($\log gf = -8.14$), and it is in reasonable agreement with the value ($\log gf 
= -8.21$) found in the Vienna Atomic Line Database 
\citep[VALD;][]{1995A&AS..112..525P,1999A&AS..138..119K,ryabch}. The atomic linelist 
for the region surrounding the [\ion{C}{1}] feature is from VALD, and it includes the 
blending \ion{Fe}{1} line at 8727.10 {\AA} ($\log gf = -3.93$).  Although we 
expect this \ion{Fe}{1} line to be of no consequence in the spectra of our 
metal-poor stars, \citet{2006MNRAS.367.1181B} determined that the line can be 
important for stars with $T_{\mathrm{eff}} < 5700 \; \mathrm{K}$, so for 
completeness, we include it in our linelist.  Transition data for three CN 
lines in the immediate vicinity of the [\ion{C}{1}] line are also included in 
the linelist and are taken from \citet{1999A&A...342..426G}.  The fits to the 
observed spectra are shown in Figure 1.

\footnotetext[8]{See http://physics.nist.gov/PhysRefData/ASD/index.html}

In addition to the $\lambda 8727$ [\ion{C}{1}] feature, we have scoured the
bHROS spectra for additional spectral lines and have identified for each star 
2--8 \ion{Fe}{1} lines, 2--3 high-excitation \ion{C}{1} permitted lines, 5--9 
resolved CN lines of the CN red system, and the \ion{K}{1} resonance line at 
7698.96 {\AA}.  The \ion{Fe}{1} lines provide a measurement of the overall 
stellar non-CNO metallicity, and our analysis provides an additional independent
determination of the Fe abundances of the stars in our sample.  Non-local 
thermodynamic equilibrium effects are expected to influence the formation of 
high-excitation \ion{C}{1} lines, so the comparison of the \ion{C}{1}-based 
abundances to those based on the [\ion{C}{1}] feature provide a empirically 
determined estimate of the magnitude of the effect in these CEMP stars.  The CN 
lines have been analyzed for N abundances, assuming the C abundance derived 
from the [\ion{C}{1}] feature.  Nitrogen abundances for CEMP stars have been 
largely derived from CN molecular bands at 3885 and 4215 {\AA} via synthetic 
synthesis \citep[e.g.,][]{2006AJ....132..137C,2007ApJ...655..492A}, so our analysis 
of individual CN lines of the red system provides an independent measurement of 
the N abundances of our sample.  We note that many additional CN lines of the 
red system are found in our spectra, but we have concentrated on a handful of 
the cleanest lines that could be securely identified.  Finally, potassium is 
predominately produced in the oxygen-burning zones of Type II supernovae 
\citep{1995ApJS..101..181W}, so determining its abundance may provide some 
insight into the chemical history of the prenatal gas from which the stars 
formed.  A sample of \ion{Fe}{1} and CN features in our bHROS spectra is shown 
in Figure 2.

Equivalent widths of the \ion{Fe}{1}, \ion{C}{1}, CN, and \ion{K}{1} lines were 
measured using the 1-D spectrum analysis software SPECTRE 
\citep{1987BAAS...19.1129F} and were then used to derive abundances with the 
{\sf abfind} driver in the MOOG package.  The atomic parameters for the 
\ion{Fe}{1}, \ion{C}{1}, and \ion{K}{1} lines were obtained from VALD.  
We made use of the extensive list of \citet{1963rspx.book.....D} in the
identification of measurable CN lines and have adopted the wavelengths therein; 
transition probabilities for the CN lines are from a proprietary list (A.
McWilliam, private communication).  The adopted atomic parameters and the 
measured EWs for each star are listed in Table 3.  In addition to EWs, 
SPECTRE was used to measure for each star the observed wavelengths of five lines 
in our linelist in order to calculate radial velocities ($V_{\mathrm{r}}$).  The 
heliocentric-corrected mean values, along with the uncertainties in the means 
($\sigma_{\mathrm{mean}} = \sigma / \sqrt{N -1}$, where $\sigma$ is the standard 
deviation and $N$ is the number of lines measured), are provided in Table 2.

\section{RESULTS \& DISCUSSION}
\label{s:analysis}
The results of the abundance derivations and error analysis are tabulated in
Table 4.  The [Fe/H] abundances, as are all relative abundances appearing
herein, are given relative to the solar abundances provided by Asplund, 
Grevesse, \& Sauval (2005a): $A_{\odot}(\mathrm{Fe}) = 
\log N_{\odot}(\mathrm{Fe}) = 7.45$, $A_{\odot}(\mathrm{C}) = 8.39$, and 
$A_{\odot}(\mathrm{N}) = 7.78$.  Before the derived abundances are discussed, 
the results of the error analysis and the comparison of the abundances derived
using the C-enhanced and solar-scaled models are presented.

\subsection{Abundance Uncertainties}
The abundance uncertainties provided in Table 4 are based on abundance 
sensitivities of each individual element to changes in the input stellar 
parameters; they do not include any systematic errors, which may be 
considerable, resulting from assumptions made in the analysis (including the 
neglect of possible 3D and NLTE effects) or other potential sources.  For the
analysis including the solar-scaled ATLAS9 models, the parameters are \teff, 
$\log \; g$, $\xi$, and [m/H], the overall stellar metallicity.  For the analysis
including the C-enhanced ATLAS12 models, in addition to the parameters 
previously listed, sensitivities to the model C and N abundances were also 
calculated.  Following \citet{2007ApJ...655..492A}, the source of the stellar 
parameters adopted for \two\ and \three, we assume uncertainties of $\pm 100 \; 
\mathrm{K}$, $\pm 0.30 \; \mathrm{dex}$, and $\pm 0.25 \; \mathrm{km \; 
s}^{-1}$ for \teff, $\log \; g$, and $\xi$, respectively.  We point out, 
however, that \citet{2007ApJ...655..492A} adopted an uncertainty of $\pm 0.30 \; 
\mathrm{km \; s}^{-1}$ for $\xi$.  Uncertainties of $\pm 0.15 \; \mathrm{dex}$ 
for [m/H], $\pm 0.20 \; \mathrm{dex}$ for C, and $\pm 0.20 \; \mathrm{dex}$ for 
N were adopted for the ATLAS12 analysis.  The final abundance uncertainties 
($\sigma_{Total}$) are the quadratic sum of the abundance uncertainties based on 
the parameter sensitivities and the uncertainty in the mean abundance for 
species whose abundance has been derived from more than one spectral line.

The abundance sensitivities to changes in the stellar parameters of the ATLAS12
models are given for HE 0054-2542 in Table 5.  The sensitivities for HE
0054-2542 are representative of those for all three stars, although some
resulting uncertainties warrant special comment.  First, as is evident in Table 
4, the abundance uncertainties arising from the ATLAS9 and ATLAS12 analyses do 
not differ greatly.  While one might assume this suggests that the derived 
abundances are not sensitive to changes in the enhanced C abundances of the 
ATLAS12 models, this is not the case for the N abundances, for which a change 
in C abundance of $\pm 0.50 \; \mathrm{dex}$ can lead to a N abundance change 
of $\pm 0.13 \; \mathrm{dex}$.  However, because the uncertainty in the N 
abundances derived from the CN lines is dominated by the sensitivities to \teff\ 
and $\log \; g$, the sensitivity to C does not greatly affect
$\sigma_{Total}(\mathrm{N})$.  Second, the $\sigma_{Total}(\mathrm{N})$ is
significantly larger than for the other elements analyzed because of the acute
sensitivity of the N abundance to \teff\ and $\log \; g$.  Its sensitivity to
model [m/H] and C abundance also contributes a small amount to 
$\sigma_{Total}(\mathrm{N})$.  Third, none of the other abundances derived for the
three stars are sensitive to the adopted N abundance, suggesting that the N
abundances do not affect significantly the structure of the ATLAS12 model
atmospheres.  Last, the $\sigma_{Total}$ for the \ion{C}{1} abundance of \two\ 
is more than a factor of 1.5 larger than those of the other two stars.  This is 
entirely due to the large uncertainty in the mean abundance, which has a 
standard deviation of 0.32 dex.  Such a large uncertainty is not entirely 
surprising given that only two \ion{C}{1} lines with vastly different line 
strengths (18 and 217 m\AA ) were measurable for this star; the sensitivity of 
its \ion{C}{1} abundance to the other parameters is comparable to those of the 
other stars.

\subsection{C-Enhanced vs. Solar-Scaled Models}
\label{s:atlas12vs9}
In Table 4, the results of both the C-enhanced ATLAS12 and solar-scaled ATLAS9 
(hereafter we will simply refer to these as ATLAS12 and ATLAS9) analyses are 
given.  The Fe, [\ion{C}{1}], and K abundances of the three stars are consistent 
between the two analyses, with abundance differences being $\leq 0.05 \; 
\mathrm{dex}$.  There are greater discordances, however, between the ATLAS12 and 
ATLAS9 abundances for \ion{C}{1} and N.  The differences are such that the 
ATLAS12 \ion{C}{1} abundances are 0.09 -- 0.14 dex lower than the ATLAS9 
abundances, whereas the ATLAS12 N abundances are consistently higher than those 
from the ATLAS9 analysis by 0.15 dex.

The differences in the \ion{C}{1} and N abundances, as well as the minor
discrepancies in the Fe, [\ion{C}{1}], and K abundances, are entirely 
attributable to the divergences in the temperature and pressure structures of 
the ATLAS12 and ATLAS9 model atmospheres. In Figures 3 -- 5, temperature is 
plotted versus the logarithm of the pressure from the ATLAS12 and ATLAS9 models 
of \one, \two, and \three, respectively.  In each figure, the optical 
depths\footnotemark[9] at reference wavelength 5000 {\AA} ($\tau_{5000}$) that 
bracket the line forming regions of the spectral features included in our 
analysis are marked; the inset included with each figure highlights the 
temperature-$\log P_{\mathrm{gas}}$ relations of these regions.  As can be seen
in the figures, the ATLAS12 model is hotter than the ATLAS9 model in the line 
forming region of each star except at the deepest boundaries for \one\ and
\three, the two most Fe-deficient stars, where the ATLAS12 model dips below the
ATLAS9 model.  Quantifying accurately the effect of the model differences on 
the abundances derived from each spectral line is difficult due to the 
convolution of the effects of the temperature and pressure differences at a 
given $\tau_{5000}$, but the differences of the models at the mean depths of 
formation for each line are such that they qualitatively account for the 
abundance discrepancies that are seen.  For example, the mean 
depths of formation of the \ion{C}{1} lines differ between the models in such a 
way that the ATLAS9 temperature at this layer is less than the ATLAS12 
temperature and conversely, the ATLAS9 pressure is greater than the ATLAS12 
pressure.  Both of these differences conspire to result in an ATLAS9 abundance 
that is greater than the ATLAS12 one, which is seen for all three stars.

\footnotetext[9]{The $\tau_{5000}$ values have been taken from the output files 
of the MOOG package and represent the optical depths of the mean depth of 
formation for each spectral line.}

{\it The differences in the ATLAS12 and ATLAS9 abundances of \ion{C}{1} and N 
demonstrate that model atmospheres characterized by metal abundances scaled 
from the solar composition are not ideally suited for chemical abundance 
analyses of CEMP stars}.  This result corroborates the conclusion of 
\citet{2000A&A...353..557H}.  The comparison of their C-enhanced and 
solar-scaled models are similar to those seen here for the two most 
Fe-deficient stars \one\ and \three: the C-enhanced model is hotter at a given 
pressure than the solar-scaled model in the deepest atmospheric layers, and it 
is cooler in the outer layers.  Hill et al. point out that very strong lines, 
such as the molecular band heads, would be stronger in the C-enhanced model 
because of the lower temperature in the outer regions in which these lines are 
expected to form, and that the intermediate and weak lines that form deeper in 
the atmospheres, near $\log \; \tau_{5000} = -1$, will be weaker in the 
C-enhanced models due to their higher temperatures in these layers.  Because of 
these structural differences, Hill et al. conclude that C-enhanced models should 
be used in analyses of CEMP stars if possible.

As stated above, accurately quantifying the effect of using solar-scaled as
opposed to C-enhanced model atmospheres in abundance analyses of CEMP stars is 
challenging.  The differences in the derived abundance for any given element 
are dependent on both the structural differences in the temperature and pressure
profiles of the model atmospheres and the depth of formation of each individual 
spectral line.  This is evident by the differing results for Fe, [\ion{C}{1}],
and K, which show no or only minor differences in the ATLAS12 and ATLAS9 
abundances, and for \ion{C}{1} and N, which show more significant 
discrepancies.  The magnitude of the effect is expected to be even more severe 
for the molecular band heads, which are highly sensitive to temperature and 
pressure, that are often used in the derivation of C, N, and O abundances of 
CEMP stars.  Any inaccuracies in the derived abundances of CEMP stars will 
negatively affect the interpretation of the nucleosynthetic history of these 
stars and their role in the chemical evolution of the Galaxy.  Under the 
assumption that C-enhanced models more accurately represent the real atmospheres 
of CEMP stars, it is clear that C-enhanced models should be used in their 
abundance analyses.  Moreover, we hope our analysis will provide sufficient 
motivation for a more rigorous examination of C-enhanced model atmospheres and 
their effect on the derived abundances of CEMP stars.   

Unless explicitly stated to the contrary, the remaining discussion will refer to
the abundances derived using the C-enhanced ATLAS12 models.

\subsection{Abundances}
\label{s:abundances}
\subsubsection{Iron Abundances}
The limited wavelength coverage of the bHROS spectra and the Fe deficiency of 
the stars in our sample combine to restrict the number of lines ($ N \leq 8$) 
from which Fe abundances can be derived for each star.  For the two most 
Fe-deficient stars, \one\ and \three, only three and two Fe lines, 
respectively, were measurable.  Regardless, the derived Fe abundances of all
three stars (Table 4) are in very good agreement with previously determined 
values.  \citet{2001AJ....122.1545P} and \citet{2002ApJ...580.1149A} have both 
analyzed \one\ and have found respectively $\mathrm{[Fe/H]} = -2.67$ and 
$\mathrm{[Fe/H]} = -2.64$; our value of $\mathrm{[Fe/H]} = -2.66 \pm 0.08$ 
falls between their results.  Our abundances of $\mathrm{[Fe/H]} = -1.42 \pm 
0.15$ for \two\ and $\mathrm{[Fe/H]} = -3.08 \pm 0.13$ for \three\ are in good 
agreement with those of \citet{2007ApJ...655..492A}, who derived 
$\mathrm{[Fe/H]} = -1.38$ for the former and $\mathrm{[Fe/H]} = -3.17$ for the 
latter.  Thus, despite utilizing different spectral lines and different model 
atmospheres, our results confirm the subsolar Fe abundances of \one, \two, and
\three.

\subsubsection{C Abundances from the [\ion{C}{1}] Line}
As shown in Figure 1, the synthetic spectra of the $\lambda 8727$ region fit 
well the observed spectra of each star.  The resulting abundances from the best
fits of the [\ion{C}{1}] features are given in Table 4.  Adopting our derived
[Fe/H] abundances and $A_{\odot}(\mathrm{C}) = 8.39$, 
the C abundances relative to Fe are listed in Table 6 and are 
$\mathrm{[C/Fe]} = +2.12 \pm 0.14$ for \one, $\mathrm{[C/Fe]} = +1.33 \pm 0.21$ 
for \two, and $\mathrm{[C/Fe]} = +2.14 \pm 0.17$ for \three, where the stated 
uncertainty is the quadratic sum of the C and Fe uncertainties given in Table 
4.  We discuss the abundances for each star in turn.

{\noindent \bf HE 0054-2542:} similar to the [Fe/H] abundance, our [C/Fe] ratio 
from the 1D LTE analysis for this star is bracketed by the abundances derived by 
\citet{2001AJ....122.1545P} and \citet{2002ApJ...580.1149A}, and it is in good
agreement with both of their values.  Preston \& Sneden found an abundance of
$\mathrm{[C/Fe]} = +2.2$ by analyzing the CH bands at wavelengths 4228 -- 4241
{\AA} and 4358 -- 4374 {\AA}.  Their derived abundance is labeled as uncertain, 
although the uncertainty is not quantified.  The abundance from Aoki et al., 
$\mathrm{[C/Fe]} = +2.0$, is derived from the (1-0) band of the C$_2$ Swan 
system at 4735 {\AA}, and the result is also given without an uncertainty 
estimate.  Typical errors in [C/Fe] abundances derived from high-resolution 
spectra range from about 0.15 to 0.25 dex \citep[e.g.,][]{2006AJ....132..137C,
2006A&A...459..125S}, and if this typical uncertainty is adopted for the Preston 
\& Sneden and Aoki et al. studies, their C abundances are in agreement.  Both of the 
previously derived abundances are within the uncertainty associated with our 
value derived from the [\ion{C}{1}] feature, and we thus conclude that the 
three measurements are consistent.

{\noindent \bf HE 0507-1653:} the [\ion{C}{1}]-based 1D LTE abundance for this star is
in excellent agreement with the value found by \citet{2007ApJ...655..492A}, the
difference between the two results being only 0.04 dex.  Aoki et al. derived an 
abundance of $\mathrm{[C/Fe]} = +1.29 \pm 0.19$ based on an analysis of the 
C$_2$ Swan (0-1) band at 5635 {\AA}.  The (0-1) band was chosen for measurement,
because, according to the authors, the CH G-band and the C$_2$ Swan (0-0) band 
at 5170 {\AA} were saturated and not suitable for the purpose.

{\noindent \bf HE 1005-1439:}  unlike for the two more Fe-abundant stars in our
sample, the 1D LTE-based C abundance derived from the [\ion{C}{1}] feature for \three\ 
differs from the previously determined measurement.  This star was analyzed by 
\citet{2007ApJ...655..492A} as part of the same study that included \two,
although the abundance for \three\ is based on the C$_2$ Swan (0-0) band at 
5170 {\AA} instead of the (0-1) band at 5635 {\AA}.  Aoki et al. derived an 
abundance of $\mathrm{[C/Fe]} = +2.48 \pm 0.20$, which is 0.34 dex, or about a 
$2\sigma$ deviation, larger than our [\ion{C}{1}]-based abundance.  
\citet{2007ApJ...655..492A} suggest that C abundances derived from C$_2$ bands 
may be systematically higher by 0.2 dex than those derived from CH bands based 
on the results of \citet{2002PASJ...54..427A}, who derived consistent C 
abundances for the CEMP star LP 625-44 from the (1-0), (0-0), and (0-1) bands 
of the Swan system but found the abundance derived using the CH band at 4323 
{\AA} to be about 0.20 dex lower.  However, the abundances derived for \one\ 
do not support the existence of such a systematic offset; indeed, the
C$_2$-based abundance derived for this star by \citet{2002ApJ...580.1149A} is 
0.2 dex {\it lower} than the CH-based abundance derived by 
\citet{2001AJ....122.1545P}, in direct contrast to the suggested systematic
offset.  Furthermore, a similar offset has not been suggested to exist between
individual lines of the C$_2$ Swan system.  On the contrary, 
\citet{2002PASJ...54..427A} find consistent results from three different C$_2$ 
band heads using the same line list as \citet{2007ApJ...655..492A}.  Thus, there
is no reason to believe that any one of the C$_2$ Swan lines is superior to any 
of the others for abundance analyses and that the 0.34 dex difference in the 
C$_2$ and [\ion{C}{1}]-based abundances derived for \three\ is due to the 
particular C$_2$ Swan line used.

Because the [\ion{C}{1}] and C$_2$-based abundances for \three\ differ at about 
the $2\sigma$ level of the internal uncertainty, the difference cannot be 
ascribed with statistical confidence to the inability of the C$_2$ Swan lines 
to reproduce reliable LTE abundances.  Instead, the abundance difference may be 
a result of other inadequacies in either of the abundance analyses, although 
the exact component of the analyses that could be responsible for such a large 
discrepancy is not readily identifiable.  As discussed above, the adopted 
linelists and the modeling of the lines do not seem to be at fault.  The stellar
parameters used by \citet{2007ApJ...655..492A} are the same as those used in our
analysis, so while the sensitivity of the abundance derived from the 
$\lambda 5170$ C$_2$ line to changes in stellar parameters may differ from that 
of the [\ion{C}{1}]-based abundance, it is unlikely the adopted stellar 
parameters can account for the 0.34 dex difference in the derived C abundance. 
Despite the moderate $\mathrm{S/N} = 65$ of the coadded spectrum for \three, the 
[\ion{C}{1}] line is well shaped and easily identifiable (Figure 1), and an 
excellent fit with the synthetic spectrum to both the continuum and the 
[\ion{C}{1}] feature is obtained.  Furthermore, to reproduce the [C/Fe] 
abundance of \three\ derived by \citet{2007ApJ...655..492A} would require an 
equivalent width that is approximately a factor of two larger than the one 
measured in our bHROS spectrum (Figure 6).  Thus, while the spectrum of \three\ 
is of slightly lower quality than those of the other two stars in our sample, 
there is no conspicuous evidence to suggest that it is a source of significant 
error in the derived C abundance. 

Having largely ruled out other possibilities, in $\S$ \ref{s:3dnlte} we 
discuss whether 3D effects could explain the different abundance results 
from the [\ion{C}{1}] and C$_2$ lines.

\subsubsection{The High-Excitation \ion{C}{1} Lines}
Carbon abundances have been derived from high-excitation ($7.48-8.77 \;
\mathrm{eV}$) \ion{C}{1} lines in the spectra of our CEMP star sample for
comparison to the [\ion{C}{1}] results.  The \ion{C}{1} lines are known to form
out of LTE, and the measurements presented here should provide an empirical
estimate of the associated NLTE effects in these stars.  The comparison of the 1D LTE
abundances derived from the [\ion{C}{1}] and \ion{C}{1} lines is made in Table 7, and
as expected from existing NLTE calculations \citep{2006A&A...458..899F}, the 
abundances based on the \ion{C}{1} lines are greater than those 
derived from the forbidden line.  Similarly, \citet{2002PASJ...54..427A} derived 
the C abundance of the CEMP star LP 625-44 from five high-excitation \ion{C}{1} 
lines and found the abundance to be 0.45 dex greater than that derived from CH 
and C$_2$ molecular features.  The stellar parameters of LP 625-44 
($\mathrm{[Fe/H]} = -2.7$, $\mathrm{[C/Fe]} = +2.25$, $T_{\mathrm{eff}} = 
5500 \; \mathrm{K}$, and $\log g = 2.5$; Aoki et al. 2002d) are similar to those 
of \one, and the \ion{C}{1} overabundances are nearly identical for the two 
stars.  In $\S$ \ref{s:3dnlte}, we present predicted NLTE abundances for 
\ion{C}{1} and the corresponding 3D effects for [\ion{C}{1}] .

\subsubsection{Nitrogen Abundances}
Numerous lines of the CN red system are present in our bHROS spectra, and we 
have identified 5 -- 9 clean, resolved lines for each star (Table 3) suitable 
for abundance analysis.  The mean abundances derived from these lines are
provided in Table 4 and correspond to relative abundances of $\mathrm{[N/Fe]} =
+1.24 \pm 0.35 \; (0.03)$ for \one, $\mathrm{[N/Fe]} = +1.20 \pm 0.34 \; 
(0.02)$ for \two, and $\mathrm{[N/Fe]} = +2.27 \pm 0.35 \; (0.09)$ for \three\ 
(Table 6).  The number given in the parentheses following each uncertainty is 
the standard deviation in the mean abundance for each star and is indicative of 
the excellent agreement in the line-by-line abundances derived from the CN red 
system features.  As mentioned above in $\S 5.1$, the large uncertainties in the 
N abundances arise from the severe sensitivity of these molecular features to 
changes in the \teff\ and $\log g$ parameters and are typical compared to those 
from other studies \citep{2006AJ....132..137C,2007ApJ...655..492A}.

Previous N abundance determinations for the three stars in our sample have all
been based on features of the CN blue system near 3880 and 4215 {\AA} and are
all at least 0.40 dex lower than the abundances presented here.  
\citet{2001AJ....122.1545P} and \citet{2002ApJ...580.1149A} both derived N
abundances for \one\ from the 3880 {\AA} CN band and found $\mathrm{[N/Fe]} =
+0.7$ and +0.3, respectively.  \citet{2002PASJ...54..933A} revisited the 
analysis of \citet{2002ApJ...580.1149A} and determined the $\lambda 3880$ CN 
band was too strong in their spectrum to provide an accurate abundance 
estimate.  Using the $\lambda 4215$ CN band, \citet{2002PASJ...54..933A} derived 
an abundance of $\mathrm{[N/Fe]} = +0.8$, a value that is in better agreement 
with that of Preston \& Sneden but lower by 0.44 dex than our derived 
abundance.  The N abundances for \two\ and \three\ have been determined by 
\citet{2007ApJ...655..492A} from an analysis of the $\lambda 4215$ CN band for 
each star.  Their reported abundances are $\mathrm{[N/Fe]} = +0.80$ for the 
former and $\mathrm{[N/Fe]} = +1.79$ for that latter; these values are 0.40 and 
0.48 dex, respectively, lower than our abundances.

Given the large uncertainties in the N abundances, the differences between our
values and those previously published cannot be said to be statistically
significant.  Nonetheless, the disagreement is striking.  About $50\%$ of the
difference for each star is attributable to our use of the C-enhanced ATLAS12 
model atmospheres, the abundances from which are a consistent 0.15 dex higher 
than those derived from the solar-scaled ATLAS9 analysis.  The N abundances from 
our ATLAS9 analysis are $\sim 0.3$ dex higher than the literature values, and
the source of this remaining difference is not so obvious.  The adoption of 
different molecular dissociation energies by different studies can lead to 
diverging abundances, but this is not the case here.  Following 
\citet{2005A&A...430..655S} and \citet{2007ApJ...670..774N}, we have adopted a
dissociation energy of $D_{0} = 7.76 \; \mathrm{eV}$ for the CN molecule, which
is essentially the same as the value of $D_{0} = 7.75 \; \mathrm{eV}$ adopted 
by \citet{2002PASJ...54..933A} and \citet{2007ApJ...655..492A}\footnotemark[10].
\citet{2005A&A...430..655S} derived the N abundances of 10 metal-poor giants 
from both the CN blue band near 3880 {\AA} and the NH band at 3360 {\AA}, and 
the CN-based abundances were found to be systematically lower by $\sim 0.40$ 
dex than the NH-based abundances.  The reason for the discrepancy was not 
apparent to the authors, but we find it to be similar to the $\sim 0.30$ dex 
difference between the CN blue bands and CN red system abundances from the 
ATLAS9 analysis seen here.  Taken together, the results of 
\citet{2005A&A...430..655S} and of our study suggest that analyses of the CN 
blue system bands underestimate the N abundances of metal-poor giants.  Because 
the CN band is used often for the derivation of N abundances is such stars, 
further investigation into the ability of this feature to provide accurate 
abundances is needed.

\footnotetext[10]{The CN dissociation energy does not appear in the published
papers of Aoki et al. and has been provided via private communication.}

Accurately determined N abundances are critical to understanding the
nucleosynthetic history of CEMP stars and to placing constraints on stellar
evolution models.  Mass-transfer from a primary companion during its asymptotic
giant branch (AGB) phase
is thought to be the source of the enhanced C and the enhancements of other 
elements in the photospheres of the majority of CEMP stars 
\citep{2005ApJ...635..349R}.  Intermediate mass AGB stars (initial masses of 
$\sim 4 - 6 \; \mathrm{M}_{\odot}$) are predicted 
to be efficient producers of primary N \citep{2004ApJ...605..425H}, and it has
been suggested that N-rich counterparts to CEMP stars, Nitrogen-Enhanced
Metal-Poor (NEMP) stars, should also be common in the Galactic halo 
\citep{2007ApJ...658.1203J}.  The initial survey of 21 CEMP stars by Johnson et
al. was carried out using ATLAS9 models with solar-scaled abundances, and their
analysis of the NH band head near 3360 {\AA} produced no identified NEMP stars. 
Given the 0.15 dex difference between the N abundances derived here using the 
ATLAS12 and ATLAS9 atmospheric models, an investigation of the effect of
CNO-enhanced model atmospheres of potential NEMP stars on derived N abundances 
is warranted.  Otherwise, {\it bona fide} NEMP stars may not be identified.

\subsubsection{Potassium Abundances}

Potassium abundances have been determined for large samples of metal-poor stars 
by a handful groups, starting with \citet{1987A&A...178..179G} and then more 
recently by \citet{2004A&A...416.1117C}, \citet{2005MNRAS.364..712Z}, and 
\citet{2006A&A...457..645Z}.  The consensus among these studies is that the LTE
[K/Fe] ratios for metal-poor stars, irrespective of evolutionary state, are 
super-solar, and they evince substantial scatter, particularly at metallicities 
of $\mathrm{[Fe/H]} \leq -1.0$.   The \ion{K}{1} resonance line at 7699 {\AA}, 
generally the only line available in the spectra of late-type stars with which 
K abundances can be derived, is highly sensitive to NLTE effects, and large 
negative corrections (-0.2 to -0.8 dex) for LTE abundances derived from this 
line have been predicted \citep{2002PASJ...54..275T}.  The magnitude of the 
corrections have been shown to be dependent on \teff\ and $\log g$ but not 
[Fe/H], and after applying the corrections to the LTE abundances, the 
super-solar [K/Fe] ratios and the large scatter remain 
\citep[e.g.,][]{2002PASJ...54..275T,2005MNRAS.364..712Z}.

The LTE K abundances derived for the three stars in our sample are given in
Table 4 and correspond to [K/Fe] relative abundances of +0.61, +0.74, and +0.50
for \one, \two, and \three, respectively.  The K abundance has been derived for 
only one other CEMP star, HE 1410-0004 \citep{2006AJ....132..137C}, a subgiant 
with $\mathrm{[Fe/H]} = -3.02$; the K abundance for this star has been 
determined to be $\mathrm{[K/Fe]} = +0.71$.  The LTE abundances of these CEMP 
stars fall squarely among the scatter in the LTE [K/Fe] abundances seen for 
C-normal metal-poor stars (e.g., Cayrel et al. 2004; Zhang \& Zhao 2005).  The 
consistent abundances derived using the C-enhanced ATLAS12 and solar-scaled 
ATLAS 9 models and the lack of a metallicity dependence on the calculated NLTE 
corrections both suggest that the NLTE effects associated with K abundances in 
CEMP stars do not differ significantly from those in C-normal stars.  While 
targeted K NLTE calculations for CEMP stars are needed, the existing NLTE 
corrections to K abundances should, to first approximation, be adequate.

\subsection{3D AND NLTE EFFECTS}
\label{s:3dnlte}
The elemental abundances presented in $\S$ \ref{s:analysis} were estimated
based on 1D model atmospheres and LTE spectral line formation.  Since these may 
be questionable assumptions especially in the low-density regimes of metal-poor 
red giant atmospheres \citep{2005ARA&A..43..481A}, it is important to investigate the
possible systematic errors introduced in the analysis by these simplifications. 
Indeed, the difference in the \three\ C abundances derived from [\ion{C}{1}] and 
molecular lines, and the systematically lower abundances derived from [\ion{C}{1}] 
compared with \ion{C}{1} for the 1D LTE analysis may signal exactly such 3D and/or NLTE effects.

Three-dimensional time-dependent hydrodynamical model atmospheres have relatively
recently started to be employed for abundance analysis of metal-poor stars 
\citep[e.g.][]{1999A&A...346L..17A,2001A&A...372..601A,2005ARA&A..43..481A,2006ApJ...644..229A,
2006ApJ...644L.121C,2007A&A...469..687C,2008A&A...480..233G}.  The main difference in 
the atmospheric structure is much lower temperatures in the line-forming region of the 
3D model compared with classical 1D model atmospheres; these temperature differences affect 
in particular abundances derived from molecular lines, neutral species, and low-excitation lines 
\citep{1999A&A...346L..17A}.  \ion{C}{1} is a majority species and hence is less affected by 
temperature variations than, for example, \ion{Li}{1} or \ion{Fe}{1}, but because the $\lambda 
8727$ [\ion{C}{1}] line originates from a relatively low-excitation level (1.26 eV), one 
would expect the 1D abundances to be slightly overestimated at the lowest metallicities 
\citep{2005ARA&A..43..481A}.

We have carried out 3D LTE line formation calculations for the [\ion{C}{1}] 8727 \AA\
transition using the same 3D hydrodynamical model atmospheres as described in
\citet{2007A&A...469..687C}, which have similar stellar parameters as our three stars
in terms of \teff , $\log g$ and [Fe/H]. The 3D models were constructed assuming scaled
solar abundances and are thus not directly applicable to CEMP stars.  For an exploratory 
study like this we nevertheless consider this a justifiable approach.  Because the 3D 
effects are estimated from a differential comparison with 1D {\sc marcs} models 
\citep{1997A&A...318..521A} computed assuming identical stellar parameters, including the
chemical composition, to first-order the effects of a C-enriched atmosphere discussed in 
$\S$ \ref{s:atlas12vs9} will cancel out. Specifically constructed 3D models appropriate for
CEMP stars would be valuable, however, for more accurate estimates of the various 3D effects.  
We also note that a comparison with {\sc marcs} models is more relevant than with Kurucz models,
because the 3D and {\sc marcs} models are built on the same microphysics in terms of opacities 
and equation-of-state.

The $\lambda 8727$ [\ion{C}{1}] 3D line formation calculations have been computed for 
metallicities of $\mathrm{[Fe/H]}=+0$, -$1$, $-2$, and $-3$, and the 3D abundance effects 
appropriate for our three stars have been estimated and applied to the 1D LTE results
described in $\S$ \ref{s:abundances}.  The 3D-based abundances are given in Table 
\ref{t:cabundances}. As expected, the 3D effects remain modest for the [\ion{C}{1}] line:
slightly positive for \two\ and $\sim -0.1$ dex for \one\ and \three, the two most metal-poor stars. 
In this respect, [\ion{C}{1}] behaves like lines from other majority species such as \ion{O}{1}
\citep{2005ARA&A..43..481A, 2007A&A...469..687C}, but the 3D effects are less severe than 
for example for the [\ion{O}{1}] 6300 \AA\ line for a given line strength because of its 
higher excitation potential.  It should be noted that the exact value of the 3D corrections 
remains somewhat uncertain since the O abundances are not yet known in these stars; if [O/Fe] 
is significantly larger than the $+0.4$ assumed here, a greater fraction of C will be tied 
up in CO. This notwithstanding, it is clear that the 3D effects on the $\lambda 8727$ 
[\ion{C}{1}] line will remain relatively small. 

\citet{2006A&A...458..899F} have investigated the line formation of neutral C atoms
in late-type stars and have provided predicted NLTE abundance corrections for the 
high-excitation \ion{C}{1} lines employed here for stars characterized by a range of 
\teff, $\log g$, and metallicities.  The NLTE abundance corrections for \one, \two, 
and \three\ are $-0.30$, $-0.45$, and $-0.30$ dex, respectively, when assuming a C 
abundance enhancement of $\mathrm{[C/Fe]} = +0.4$ and neglecting the highly uncertain 
inelastic H collisions \citep{2005ARA&A..43..481A}.  Thus, NLTE effects on the \ion{C}{1} 
lines can explain most of the differences between the abundances derived from the 
\ion{C}{1} and [\ion{C}{1}] lines.  We note that Fabbian et al. indeed predict that 
the magnitude of the \ion{C}{1} NLTE corrections should become larger with increasing 
[C/Fe], because the line-formation is then shifted outwards in the atmosphere where 
the departures from LTE are more pronounced.  This effect has not been investigated 
computationally for C enhancements of $\mathrm{[C/Fe]} > 0.4$, and our result, combined 
with that of \citet{2002PASJ...54..427A}, suggests that such a study is very much needed.
Additionally, the 3D NLTE line formation for \ion{C}{1} has not yet been studied due to 
computational challenges arising from the large number of atomic levels needed to
be considered, but given the very substantial 1D NLTE effects uncovered here, such
calculations would also be very welcome.

As is clear from Table \ref{t:cabundances}, taking 3D and NLTE effects into account for
[\ion{C}{1}] and  \ion{C}{1}, respectively, have brought the two abundance indicators
into better agreement, from a difference of $\sim 0.5$ dex to $\sim 0.25$ dex. The 
remaining difference can be plausibly attributed to underestimated 3D and NLTE effects, 
erroneous \teff, or simply observational uncertainties.  Of the two abundance indicators,
we consider the [\ion{C}{1}]-based abundances to be the most reliable. 

As discussed in $\S$ \ref{s:abundances}, the abundances of \one\ and \two\ derived 
from the [\ion{C}{1}] line are in good agreement with those derived from CH and C$_2$
features.  \citet{2007A&A...469..687C} have shown that, assuming scaled-solar abundances,
the 3D corrections for CH are relatively constant at $\sim -0.1$ to $\sim -0.2$ dex down 
to [Fe/H]$=-2$ for giants; C$_2$ lines were not investigated.  These 3D effects would 
therefore not be expected to ruin the overall agreement for the two CEMP stars.  More 
noteworthy, the C$_2$-based abundance for \three\ from \citet{2007ApJ...655..492A} 
is $0.34$ dex larger than our 1D result for [\ion{C}{1}] and $0.49$ dex larger than the 
corresponding 3D value.  While no 3D calculations for C$_2$ have been performed to date, 
at least qualitatively this is in line with our expectations that the 3D corrections 
for molecular lines become more severe towards lower metallicities 
\citep{2001A&A...372..601A,2005ARA&A..43..481A,2007A&A...469..687C}.  Unfortunately, we 
are unable here to carry out such a study without a better knowledge of the stellar O 
abundances; the cool outer atmospheric layers of the low-metallicity 3D models
that give rise to the C$_2$ lines are particularly conducive to CO molecule formation,
which locks up a significant fraction of C and thereby limits the amount available for C$_2$
(the [\ion{C}{1}] lines will be less affected by the CO formation since they form in deeper
atmospheric layers where the temperatures are significantly higher).

\section{CONCLUSIONS}
An investigation into the accuracy of published C abundances as derived from
the traditionally used CH and C$_2$ Swan bands of CEMP stars has been presented.  
We have analyzed the $\lambda 8727$ [\ion{C}{1}] line, a normally reliable C 
abundance indicator that is impervious to NLTE effects, in high-quality 
Gemini-S/bHROS spectra of three CEMP stars and have compared the C abundances 
derived from this line to previously published values.  The 1D LTE abundances 
derived from the [\ion{C}{1}] feature are found to confirm the abundances derived 
from both CH and C$_2$ molecular features in the two most Fe-abundant stars, 
\one\ and \two\ ($[\mathrm{Fe/H}] = -2.66$ and -1.42, respectively), but the 
[\ion{C}{1}]-based abundance of the most Fe-deficient star, \three\ 
($[\mathrm{Fe/H}] = -3.08$), is 0.34 dex, or about a $2\sigma$ deviation, 
lower than the abundance derived from the C$_2$ Swan (0-0) band at 5170 {\AA}. 

As described in $\S 5.3.2$, the specific C$_2$ Swan band used to derive C
abundances does not appear to be a factor in the difference seen for \three. 
Also, no particular component of the two abundance analyses could be identified 
as the most likely source of the discordance, although internal errors may yet 
be the underlying cause.  Unidentified systematic errors in the analyses also 
cannot be ruled out as the source of the difference, but the agreement in the 
abundances of \two\ derived by us and \citet{2007ApJ...655..492A}, the study 
that also derived the discordant abundance of \three, suggests that no such 
systematic errors are to be expected.  Thus, the difference in the C abundance 
seen between the two studies may indeed be the result of 3D effects in the 
analysis of the C$_2$ Swan band.  \citet{2007A&A...469..687C} investigated the 
impact of 3D effects on CNO abundances of red giants derived from weak CH, NH, 
and OH lines and found the 3D C abundances to be 0.5 to 0.8 dex lower (for 
stars with $\mathrm{[Fe/H]} = -3.0$) than those derived using traditional 1D 
analyses.  While these corrections are larger than the difference between the 
C$_2$ and [\ion{C}{1}]-based abundances for \three, it must be pointed out that 
Collet et al. considered stars with {\it scaled-solar} metallicities, and the 
corrections may not be applicable to CEMP stars.  Also, the 3D corrections for 
CH lines are not necessarily similar to those for C$_2$ lines (e.g., Asplund et 
al. 2005b; Collet et al. 2006).

To investigate the potential 3D effects on [\ion{C}{1}]-based abundances of 
low-metallicity stars, we have carried out 3D LTE line formation calculations 
for [\ion{C}{1}] using the same 3D hydrodynamical model atmospheres as described 
by \citet{2007A&A...469..687C}.  The 3D correction for \two\ is estimated to be 
slightly positive (+0.03 dex), while the corrections for \one\ and \three\ are 
negative (-0.07 and -0.15 dex, respectively).  Thus, the corrections are quite 
modest but do increase in magnitude at lower metallicities, similar to what is seen 
for the [\ion{O}{1}] line \citep{2005ARA&A..43..481A,2007A&A...469..687C}.  While 
more accurate [\ion{C}{1}] 3D corrections for these CEMP stars await C-enhanced 3D
models and more certain O abundances, the observations and calculations presented 
here are qualitatively in line with expectations that the 3D corrections for molecular
line-based C abundances become more severe towards lower metallicities.  

We have also determined abundances from high-excitation \ion{C}{1} lines and found
them to be systematically higher than the [\ion{C}{1}]-based abundances by 
$\sim 0.5$ dex when assuming LTE. One-dimensional abundance corrections have been 
taken from \citet{2006A&A...458..899F}, who performed NLTE calculations for a range 
of stellar parameters assuming a C abundance enhancement of $\mathrm{[C/Fe]} = +0.4$, 
and the NLTE-corrected \ion{C}{1} abundances are in significantly better agreement 
with the 3D [\ion{C}{1}]-based values.  However, there remains a $\sim 0.25$ dex 
difference in the two C abundances which can be tentatively attributed to 
underestimated NLTE effects on the \ion{C}{1} line because calculations for 
$\mathrm{[C/Fe]}>+0.4$, which are appropriate for our sample, have not been investigated, 
or possibly erroneous \teff\ values.

At this time, the complicated 3D and NLTE effects that may impact [\ion{C}{1}], 
\ion{C}{1}, CH, and C$_2$ features in the spectra of CEMP stars have not yet been 
fully established.  Analyses of the [\ion{C}{1}] line in high-resolution spectra of 
additional CEMP stars, particularly those at the lowest metallicities, for which CH 
and/or C$_2$-based abundances exist or are forthcoming are needed to determine if 
abundances derived from the forbidden line diverge in a systematic way from those 
derived from the molecular features.  Additionally, 3D models and NLTE line formation 
calculations with chemical compositions appropriate for CEMP stars are clearly needed 
to complement the observations.  These measures are necessary so that the most accurate 
C abundances possible can be derived for CEMP stars.

\acknowledgements
Support for S.C.S. has been provided by the NOAO Leo Goldberg Fellowship; NOAO
is operated by the Association of Universities for Research Astronomy (AURA), 
Inc., under a cooperative agreement with the National Science Foundation.  K.C. 
and V.V.S. acknowledge support from the National Science Foundation (NSF) under 
grant AST 06-46790.  T.C.B. and T.S. acknowledge partial support for this
work from the NSF under grants AST 04-06784, AST 07-07776, and PHY 02-16783; 
Physics Frontier Center/Joint Institute for Nuclear Astrophysics (JINA).  We 
thank W. Aoki for providing the CN molecule dissociation energy that was used 
in his studies.  This paper is based on observations obtained with the bHROS 
optical spectrograph and the Gemini Observatory, which is operated by AURA, 
Inc., under a cooperative agreement with the NSF on behalf of the Gemini 
partnership: the National Science Foundation (United States), the Science and 
Technology Facilities Council (United Kingdom), the National Research Council 
(Canada), CONICYT (Chile), the Australian Research Council (Australia), 
Ministério da Ciência e Tecnologia (Brazil) and SECYT (Argentina).  The 
observations were conducted under Gemini program GS-2006B-DD-96.

{\it Facility: } \facility{Gemini:South (bHROS)}


\begin{figure}
\plotone{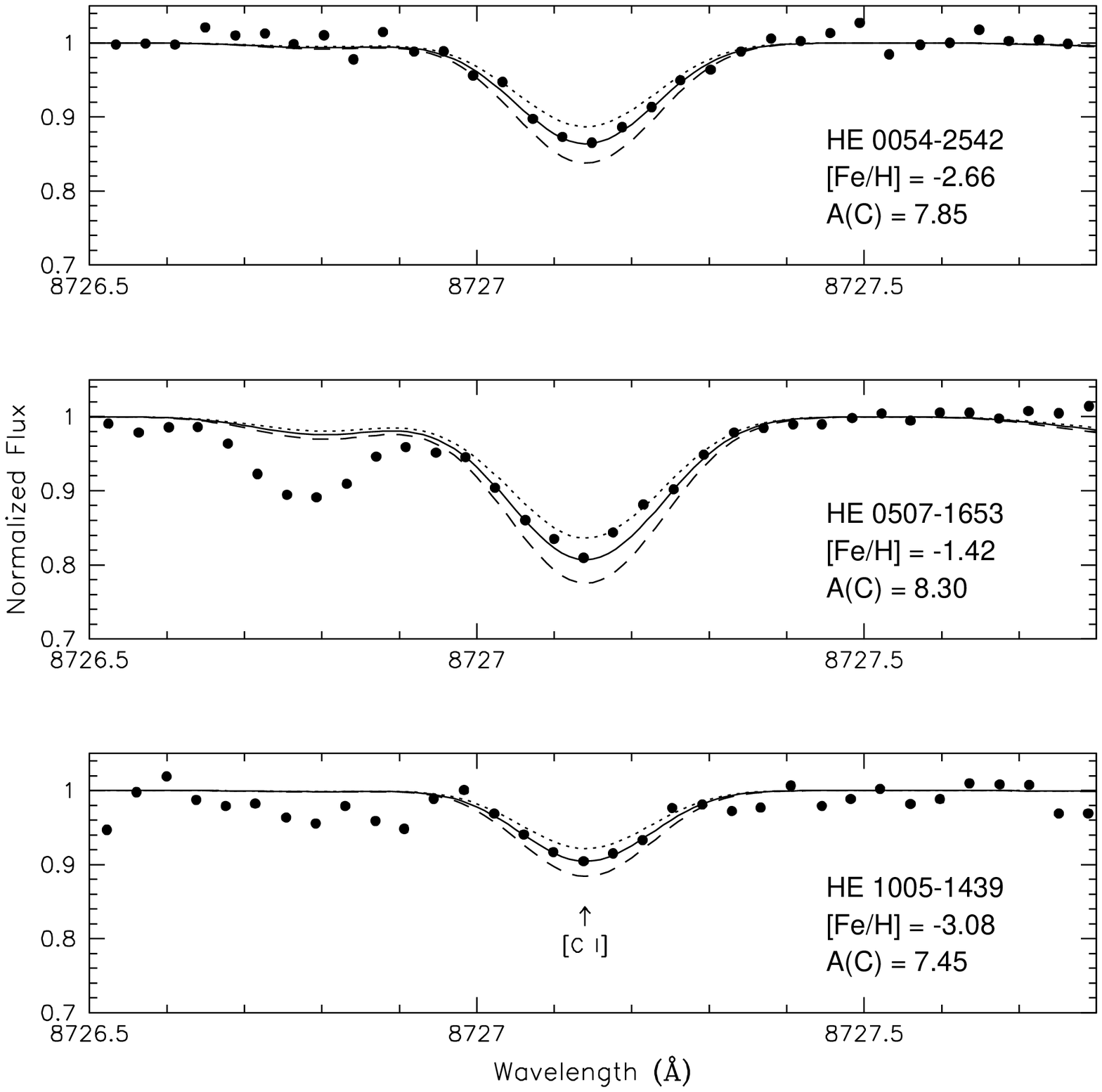}
\caption{High-resolution bHROS spectra (filled circles) and synthetic fits of 
the 8727 {\AA} region.  The solid line is the best-fit synthetic spectrum, 
characterized by the 1D LTE C abundance given in each panel.  The broken lines 
represent $\pm 0.10$ dex the best-fit abundance.  The [\ion{C}{1}] feature at 
8727.13 {\AA} is marked by the arrow in the bottom panel.}
\end{figure}

\begin{figure}
\plotone{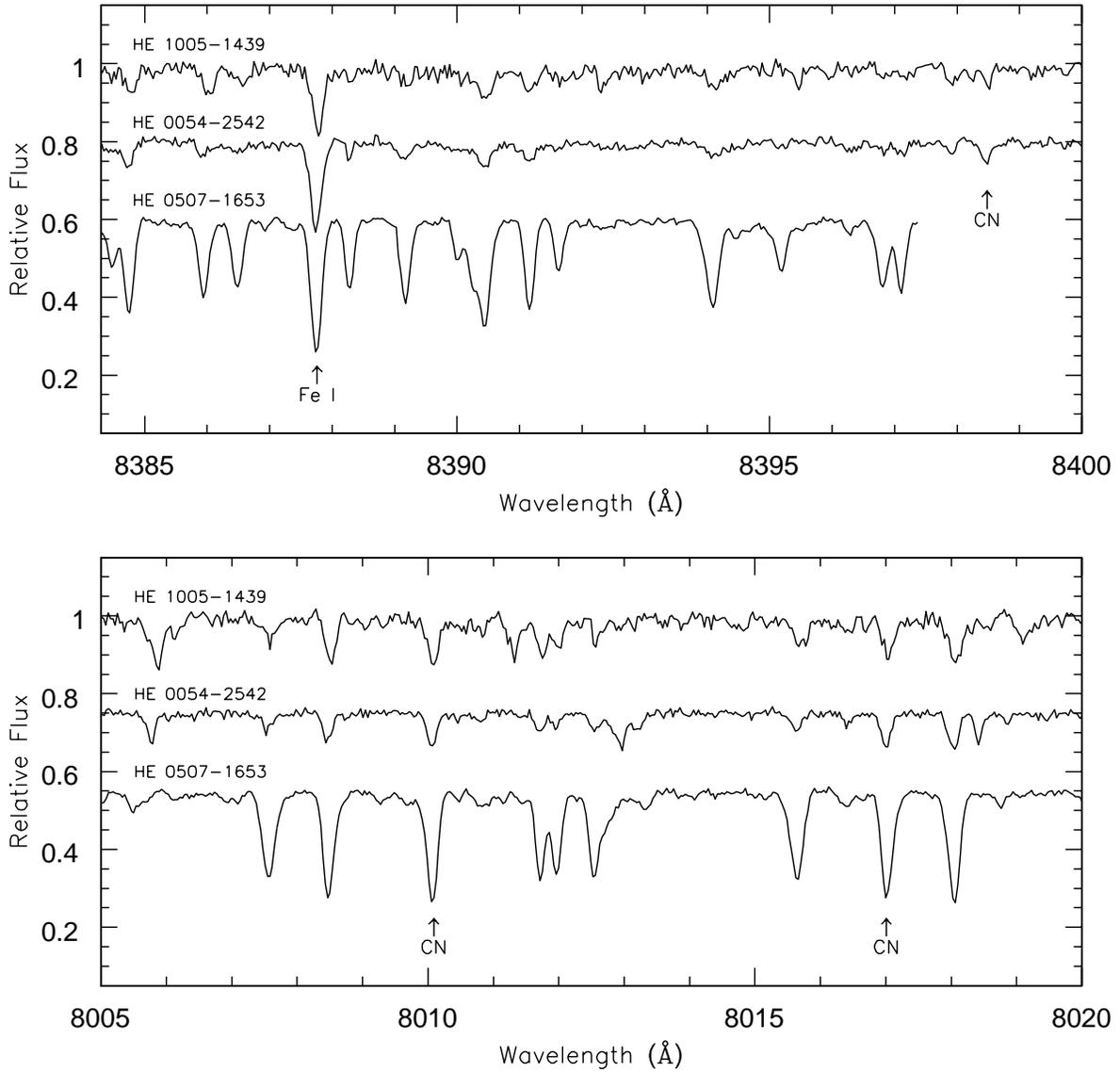}
\caption{High-resolution bHROS spectra of our stellar sample.  The lines that 
have been measured and that have been utilized in the abundance analysis are 
marked.  All of the marked lines have been measured for each star except the CN 
line at 8398.48 {\AA}, which was measurable only in the spectrum of HE 
0054-2542.}
\end{figure}

\begin{figure}
\plotone{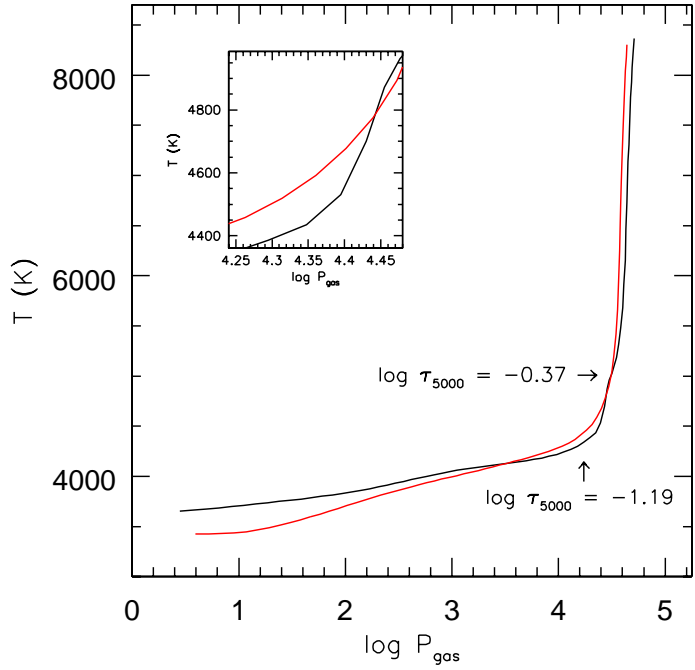}
\caption{Temperature vs. the logarithm of the gas pressure of the1D LTE C-enhanced 
ATLAS12 (red) and solar-scaled ATLAS9 (black) models for HE 0054-2542.  The 
inset shows the relations in the line forming region of the spectral features 
considered in our analysis, as marked by the upper and lower optical depths 
($\log \tau_{5000}$) in the main panel.}
\end{figure}

\begin{figure}
\plotone{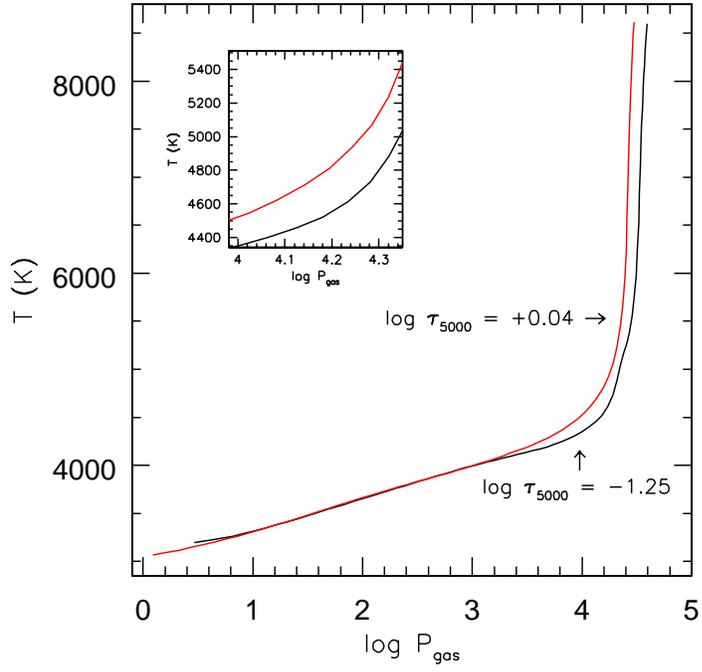}
\caption{The same as Figure 3 but for \two.}
\end{figure}

\begin{figure}
\plotone{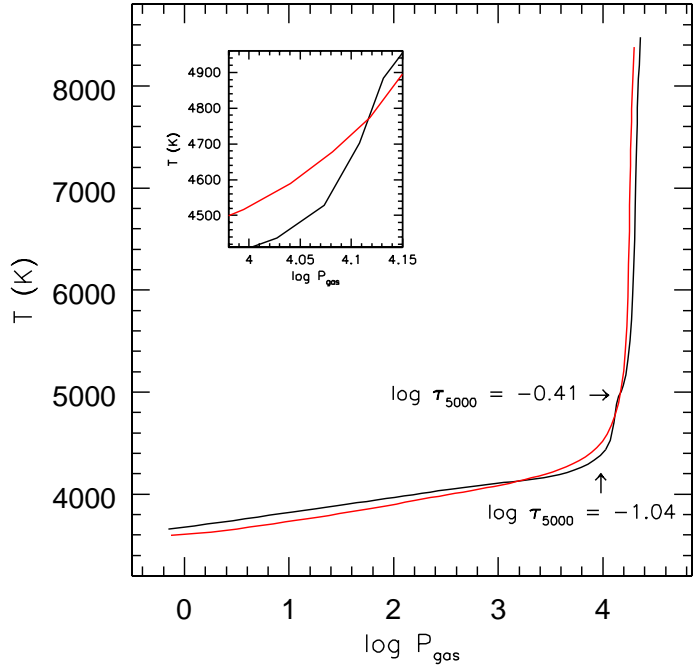}
\caption{The same as Figure 3 but for \three.}
\end{figure}

\begin{figure}
\plotone{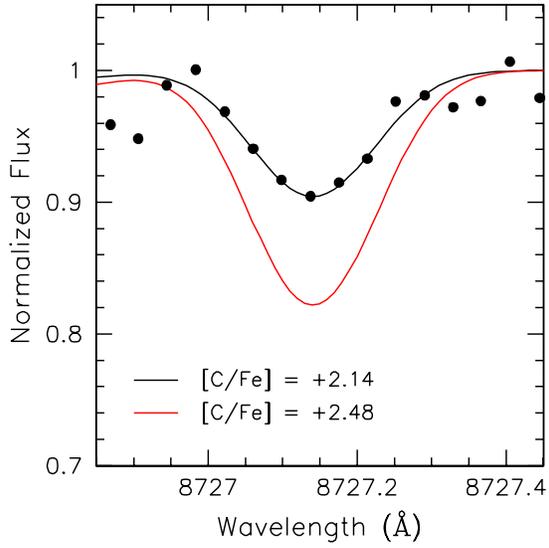}
\caption{Synthetic spectra of the [\ion{C}{1}] line of \three.  The observed
spectrum is given as filled circles, and the best fit synthetic spectrum from 
our analysis, corresponding to an 1D LTE abundance of $\mathrm{[C/Fe]} = +2.14$, is
shown as the solid black line.  The solid red line represents the synthesis of
the [\ion{C}{1}] feature assuming an abundance of $\mathrm{[C/Fe]} = +2.48$, 
the C abundance derived for this star by \citet{2007ApJ...655..492A}.  The
equivalent width of the synthetic line characterized by the higher abundance is
approximately a factor of two larger than that based on our result.}
\end{figure}

\input{tab1.tex}
\input{tab2.tex}
\input{tab3.tex}
\input{tab4.tex}
\input{tab5.tex}
\input{tab6.tex}
\input{tab7.tex}

\end{document}

%% file: tab1.tex
\begin{deluxetable}{lrcrcrc}
\tablecolumns{7}
\tablewidth{0pt}
\tablecaption{Observing Log}
\tablehead{
     \colhead{}&
     \colhead{}&
     \colhead{}&
     \colhead{}&
     \colhead{}&
     \colhead{}&
     \colhead{Integration Time}\\
     \colhead{Star}&
     \colhead{}&
     \colhead{UT Date}&
     \colhead{}&
     \colhead{Exposures}&
     \colhead{}&
     \colhead{(s)}
     }

\startdata
HE 0054-2542 && 2006 Dec 25 && 1 && 3140\\
             && 2006 Dec 25 && 1 && 3600\\
	     && 2006 Dec 26 && 1 && 3140\\
HE 0507-1653 && 2006 Dec 25 && 2 && 3140\\
HE 1005-1439 && 2006 Dec 25 && 3 && 3400\\
             && 2006 Dec 25 && 1 && 2400\\
	     && 2006 Dec 26 && 1 && 3300\\
\enddata

\end{deluxetable}

%% file: tab2.tex
\begin{deluxetable}{lccccccccccccccccrrc}
\tablecolumns{20}
\rotate
\tabletypesize{\footnotesize}
\tablewidth{0pt}
\tablecaption{Stellar Parameters}
\tablehead{
     \colhead{}&
     \colhead{}&
     \multicolumn{6}{c}{Photometry}&
     \colhead{}&
     \colhead{}&
     \colhead{$T_{\mathrm{eff}}$}&
     \colhead{}&
     \colhead{$\log g$}&
     \colhead{$\xi$}&
     \colhead{}&
     \colhead{}&
     \colhead{}&
     \colhead{}&
     \colhead{$V_{\mathrm{r}}$}&
     \colhead{$\sigma_{\mathrm{mean}}$}\\
     \cline{3-8}\\
     \colhead{Star}&
     \colhead{}&
     \colhead{$B$}&
     \colhead{$V$}&
     \colhead{$R$}&
     \colhead{$I$}&
     \colhead{$J$}&
     \colhead{$K$}&
     \colhead{}&
     \colhead{}&
     \colhead{(K)}&
     \colhead{}&
     \colhead{(cgs)}&
     \colhead{(km s$^{-1}$)}&
     \colhead{[Fe/H]}&
     \colhead{[C/Fe]}&
     \colhead{[N/Fe]}&
     \colhead{}&
     \colhead{(km s$^{-1}$)}&
     \colhead{(km s$^{-1}$)}
     }

\startdata
HE 0054-2542 && 13.57 & 12.71 &\nodata&\nodata& 11.19 & 10.65 &&& 5000 && 2.40 & 2.00 & -2.64 & +2.00 & +0.80 && -230.84 & 0.44\\
HE 0507-1653 && 13.63 & 12.51 & 12.00 & 11.58 & 10.88 & 10.32 &&& 5000 && 2.40 & 2.00 & -1.38 & +1.30 & +0.80 &&  358.94 & 0.25\\
HE 1005-1439 && 14.44 & 13.52 & 13.05 & 12.52 & 12.24 & 11.70 &&& 5000 && 1.90 & 2.00 & -3.17 & +2.50 & +1.80 &&   87.65 & 0.25\\
\enddata

\tablecomments{$B$ and $V$ magnitudes for HE 0054-2542 are from Aoki et al. 2002c; $B,V,R,$ and $I$ 
magnitudes for HE 0507-1653 and HE 1005-1439 are from Beers et al. 2007.  $J$ and $K$ magnitudes for 
all three stars are from the 2MASS Point Source Catalog (Skrutskie et al. 2006).}

\end{deluxetable}

%% file: tab3.tex
\begin{deluxetable}{lccccccrcrcr}
\tablecolumns{12}
\tabletypesize{\small}
\tablewidth{0pt}
\tablecaption{Equivalent Width Measurements}
\tablehead{
     \colhead{}&
     \colhead{$\lambda_{\mathrm{rest}}$}&
     \colhead{}&
     \colhead{$\chi$}&
     \colhead{}&
     \colhead{}&
     \colhead{}&
     \multicolumn{5}{c}{EWs (m{\AA})}\\
     \cline{8-12}\\
     \colhead{Species}&
     \colhead{({\AA})}&
     \colhead{}&
     \colhead{(eV)}&
     \colhead{}&
     \colhead{$\log gf$}&
     \colhead{}&
     \colhead{HE 0054-2542}&
     \colhead{}&
     \colhead{HE 0507-1653}&
     \colhead{}&
     \colhead{HE 1005-1439}
     }
     
\startdata
\ion{Fe}{1}\ldots & 7181.19 && 4.22 && -0.884 &&\nodata&&  24.5 &&\nodata\\ 
                  & 7445.76 && 4.26 && -0.237 &&   9.6 &&  74.5 &&\nodata\\
		  & 7723.21 && 2.28 && -3.617 &&\nodata&&  19.3 &&\nodata\\  
		  & 7998.94 && 4.37 &&  0.048 &&\nodata&&  79.9 &&\nodata\\
		  & 8046.05 && 4.42 && -0.082 &&  10.1 &&  73.8 &&\nodata\\
		  & 8327.06 && 2.20 && -1.525 &&\nodata&& 147.3 &&\nodata\\
		  & 8387.78 && 2.18 && -1.493 &&  59.4 && 126.1 && 38.2  \\
		  & 8688.64 && 2.18 && -1.212 &&\nodata&&\nodata&& 49.8  \\
		  & 8699.45 && 4.96 && -0.380 &&\nodata&&  25.6 &&\nodata\\
\ion{C}{1}\ldots  & 7685.20 && 8.77 && -1.519 &&\nodata&&  18.3 &&\nodata\\
                  & 8335.15 && 7.69 && -0.420 &&\nodata&&\nodata&& 78.8  \\
                  & 9088.51 && 7.48 && -0.429 && 130.0 &&\nodata&&115.4  \\
		  & 9094.83 && 7.49 &&  0.150 && 185.2 && 216.8 &&147.4  \\
CN\ldots          & 7185.15 && 0.42 && -1.604 &&   9.1 &&\nodata&& 12.1  \\
                  & 8010.09 && 0.19 && -1.480 &&  20.9 && 107.0 && 24.9  \\
		  & 8017.01 && 0.22 && -1.463 &&  21.7 && 106.4 && 25.7  \\
		  & 8053.09 && 0.26 && -1.406 &&  18.7 && 107.5 && 22.0  \\
		  & 8054.09 && 0.28 && -1.406 &&  21.5 && 107.6 && 18.6  \\
		  & 8064.11 && 0.29 && -1.393 &&  20.7 &&\nodata&& 20.3  \\
		  & 8074.43 && 0.31 && -1.380 &&  21.2 &&\nodata&&\nodata\\
		  & 8345.73 && 0.65 && -1.139 &&  13.8 &&  87.9 && 23.0  \\
		  & 8398.48 && 0.73 && -1.100 &&  14.0 &&\nodata&&\nodata\\
\ion{K}{1}\ldots  & 7698.96 && 0.00 && -0.170 &&  35.9 && 131.8 && 13.7  \\
\enddata

\end{deluxetable}

%% file: tab4.tex
\begin{deluxetable}{lrrrrrrrr}
\tablecolumns{9}
\rotate
\tablewidth{0pt}
\tablecaption{Derived Abundances and Uncertainties}
\tablehead{
     \colhead{}&
     \colhead{}&
     \multicolumn{3}{c}{ATLAS12: C-Enhanced Models}&
     \colhead{}&
     \multicolumn{3}{c}{ATLAS9: Solar-Scaled Models}\\
     \cline{3-5} \cline{7-9}\\
     \colhead{Species}&
     \colhead{}&
     \colhead{HE 0054-2542}&
     \colhead{HE 0507-1653}&
     \colhead{HE 1005-1439}&
     \colhead{}&
     \colhead{HE 0054-2542}&
     \colhead{HE 0507-1653}&
     \colhead{HE 1005-1439}
     }
     
\startdata
[Fe/H]\tablenotemark{a}\dotfill && -2.66 & -1.42 & -3.08 && -2.63 & -1.42 & -3.09\\
$\sigma$\dotfill                &&  0.08 &  0.15 &  0.13 &&  0.10 &  0.16 &  0.14\\
$A$([\ion{C}{1}])\dotfill       &&  7.85 &  8.30 &  7.45 &&  7.85 &  8.26 &  7.45\\
$\sigma$\dotfill                &&  0.12 &  0.14 &  0.11 &&  0.11 &  0.13 &  0.12\\
$A$(\ion{C}{1})\dotfill         &&  8.34 &  8.94 &  7.90 &&  8.43 &  9.08 &  7.99\\
$\sigma$\dotfill                &&  0.08 &  0.34 &  0.15 &&  0.12 &  0.42 &  0.17\\
$A$(N)\dotfill                  &&  6.36 &  7.56 &  6.97 &&  6.21 &  7.41 &  6.82\\
$\sigma$\dotfill                &&  0.34 &  0.31 &  0.33 &&  0.36 &  0.32 &  0.37\\
$A$(K)\dotfill                  &&  3.03 &  4.40 &  2.50 &&  3.05 &  4.45 &  2.53\\
$\sigma$\dotfill                &&  0.08 &  0.19 &  0.07 &&  0.09 &  0.21 &  0.08\\
\enddata

\tablenotetext{a}{Calculated using the solar Fe abundance of Asplund et al. (2005a)}

\end{deluxetable}

%% file: tab5.tex
\begin{deluxetable}{lcccccccc}
\tablecolumns{9}
\rotate
\tablewidth{0pt}
\tablecaption{Representative Abundance Sensitiviteis\tablenotemark{a}}
\tablehead{
     \colhead{}&
     \colhead{}&
     \colhead{$\Delta T$}&
     \colhead{$\Delta \log \; g$}&
     \colhead{$\Delta \xi$}&
     \colhead{$\Delta \mathrm{[m/H]}$}&
     \colhead{$\Delta A(\mathrm{C})$}&
     \colhead{$\Delta A(\mathrm{N})$}&
     \colhead{}\\
     \colhead{Species}&
     \colhead{}&
     \colhead{($\pm 150 \; \mathrm{K}$)}&
     \colhead{($\pm 0.50 \; \mathrm{dex}$)}&
     \colhead{($\pm 0.25 \; \mathrm{km s}^{-1}$)}&
     \colhead{($\pm 0.50 \; \mathrm{dex}$)}&
     \colhead{($\pm 0.50 \; \mathrm{dex}$)}&
     \colhead{($\pm 0.50 \; \mathrm{dex}$)}&
     \colhead{$\sigma_{Total}$\tablenotemark{b}}
     }
     
\startdata
Fe              \dotfill && $\pm 0.11$ & $0.00$              & $\pm 0.02$ & $\pm 0.04$         & $\pm 0.06$ & $0.00$ & $\pm 0.08$\\
$[$\ion{C}{1}$]$\dotfill && $\pm 0.07$ & $\pm 0.18$          & $0.00$     & $^{+0.07}_{-0.04}$ & $\pm 0.04$ & $0.00$ & $\pm 0.12$\\
\ion{C}{1}      \dotfill && $\mp 0.08$ & $^{+0.07}_{-0.10}$  & $\mp 0.05$ & $0.00$             & $\mp 0.02$ & $0.00$ & $\pm 0.08$\\
N               \dotfill && $\pm 0.38$ & $\pm 0.33$          & $0.00$     & $^{+0.15}_{-0.10}$ & $\pm 0.13$ & $0.00$ & $\pm 0.34$\\
K               \dotfill && $\pm 0.11$ & $0.00$              & $\pm 0.02$ & $\pm 0.04$         & $\pm 0.05$ & $0.00$ & $\pm 0.08$\\
\enddata

\tablenotetext{a}{Abundance senstivities are given for HE 0054-2542}
\tablenotetext{b}{The total abundance uncertainty, given by the quadratic sum 
                  of the individual parameter sensitivities and the uncertainty
		  in the mean abundance, for abundances based on more than a
		  single spectral line.}

\end{deluxetable}

%% file: tab6.tex
\begin{deluxetable}{lcccccccc}
\tablecolumns{9}
\tablewidth{0pt}
\tablecaption{Carbon and Nitrogen Abundances}
\tablehead{
     \colhead{Star}&
     \colhead{}&
     \colhead{$A(\mathrm{C})$}&
     \colhead{}&
     \colhead{[C/Fe]}&
     \colhead{}&
     \colhead{$A(\mathrm{N})$}&
     \colhead{}&
     \colhead{[N/Fe]}
     }
     
\startdata
HE 0054-2542 && 7.85 && +2.12 && 6.36 && +1.24\\
HE 0507-1653 && 8.30 && +1.33 && 7.56 && +1.20\\
HE 1005-1439 && 7.45 && +2.14 && 6.97 && +2.27\\
\enddata

\end{deluxetable}

%% file: tab7.tex
\begin{deluxetable}{lccccccc}
\tablecolumns{8}
\tablewidth{0pt}
\tablecaption{Carbon Abundance Comparison}
\label{t:cabundances}
\tablehead{
     \colhead{Star}&
     \colhead{[Fe/H]}&
     \colhead{$A$([\ion{C}{1}])}&
     \colhead{$A$([\ion{C}{1}])}&
     \colhead{$A$(\ion{C}{1})}&
     \colhead{$A$(\ion{C}{1})}&
     \colhead{$\Delta A$(C)\tablenotemark{a}} &
     \colhead{$\Delta A$(C)\tablenotemark{a}} \\
     \colhead{}&
     \colhead{}&
     \colhead{1D LTE}&
     \colhead{3D LTE}&
     \colhead{1D LTE}&
     \colhead{1D NLTE}&
     \colhead{1D LTE}&
     \colhead{3D + NLTE}

     }
     
\startdata
HE 0054-2542 & -2.66 & 7.85 & 7.78 & 8.34 & 8.04 & +0.49 & +0.26 \\
HE 0507-1653 & -1.42 & 8.30 & 8.33 & 8.94 & 8.49 & +0.64 & +0.16 \\
HE 1005-1439 & -3.08 & 7.45 & 7.30 & 7.90 & 7.60 & +0.45 & +0.30 \\
\enddata

\tablenotetext{a}{$\Delta A$(C) $= A$(\ion{C}{1})$- A$([\ion{C}{1}])}

\end{deluxetable}